\pdfoutput=1
\documentclass[a4paper, 11pt, oneside]{article}

\usepackage{epsf,amsmath,amssymb,graphicx,dcolumn,slashed}
\usepackage{scalefnt,cite,hyperref,bbold}
\usepackage{cleveref,etoolbox,color,soul}
\usepackage{listings,booktabs}
\usepackage{epstopdf}
\usepackage{amsfonts}
\usepackage{xcolor}

\usepackage{comment}

\newlength{\absize}

\newcommand{\tb}{\tan\beta}
\newcommand{\cba}{\cos(\beta - \alpha)}
\newcommand{\mA}{m_A}
\newcommand{\tev}{\,\, \mathrm{TeV}}
\newcommand{\gev}{\,\, \mathrm{GeV}}
\newcommand{\anf}{\ensuremath{A_{95}}}

\newcommand{\beq}{\begin{equation}}  
\newcommand{\eeq}{\end{equation}}
\newcommand{\bea}{\begin{eqnarray}} 
\newcommand{\eea}{\end{eqnarray}}

\newcommand\refeq[1]{Eq.~(\ref{#1})}

\newcommand\refta[1]{Tab.~\ref{#1}}
\newcommand\reftas[1]{Tabs.~\ref{#1}}
\newcommand\refse[1]{Sect.~\ref{#1}}

\newcommand\citere[1]{Ref.~\cite{#1}}
\newcommand\citeres[1]{Refs.~\cite{#1}}

\def\reffi#1{\mbox{Fig.~\ref{#1}}}
\def\reffis#1{\mbox{Figs.~\ref{#1}}}

\definecolor{Orange}{named}{orange}
\definecolor{Red}{named}{red}
\definecolor{Blue}{named}{blue}
\definecolor{Green}{named}{green}

\newcounter{notecount}
\begin{document}

\long\def\symbolfootnote[#1]#2{\begingroup%
\def\thefootnote{\fnsymbol{footnote}}\footnote[#1]{#2}\endgroup}

\title{
\Large\bf\boldmath
Pseudoscalar contributions to $Zh$ production at the LHC\\[.5em]
at 95 GeV and above
\unboldmath
}

\author{
  \addtocounter{footnote}{2}
  J.~Dutta$^{(1,2,3)}$\thanks{\tt juhidutta@phys.ntu.edu.tw}\,,   
  P.~M.~Ferreira$^{(4,5)}$\thanks{\tt pmmferreira@fc.ul.pt}~~and
  S.~Heinemeyer$^{(6)}$\thanks{\tt Sven.Heinemeyer@cern.ch}
  \\*[3mm]
  $^{(1)}\!$
  \small The Institute of Mathematical Sciences, \\ \small 4th Cross Street, CIT Campus, Tharamani, Chennai  600113, Tamil Nadu, India
  \\[2mm]
   $^{(2)}\!\!$
  \small Homi Bhabha National Institute,  \\
  \small Training School Complex, Anushakti Nagar, Mumbai 400094, Maharashtra, India 
  \\[2mm]
   $^{(3)}\!$
  \small Department of Physics, National Taiwan University, Taipei 10617, Taiwan. 
   \\[2mm]
 $^{(4)}\!$
   \small Instituto~Superior~de~Engenharia~de~Lisboa~---~ISEL,
  1959-007~Lisboa, Portugal
  \\[2mm]
  $^{(5)}\!$
  \small Centro~de~F\'\i sica~Te\'orica~e~Computacional,
  Faculdade~de~Ci\^encias, Universidade~de~Lisboa, \\
  \small Av.~Prof.~Gama~Pinto~2, 1649-003~Lisboa, Portugal
  \\[2mm]
  $^{(6)}\!$
  \small Instituto de Física Teórica UAM-CSIC, Cantoblanco, 28049, Madrid, Spain
  \\[2mm]
}


\maketitle

\begin{abstract}
\noindent
In generalizations of the Standard Models with extended scalar sectors, pseudoscalar particles will
contribute to associated production of 
the Higgs boson discovered at the LHC, $h$, and the $Z$ boson.
The pseudoscalar can be produced 
via gluon-gluon fusion, and as such may give contributions to $Zh$ production comparable to the 
value predicted by the Standard Model.
We analyze this possibility in two different models, the Two Higgs Doublet Model, with and without
an added complex singlet scalar, computing both the total and differential cross sections for this process. 
We focus on two pseudoscalar mass regions: (i) $m_A \sim 95 \gev$, to describe the di-photon excesses 
observed by CMS and ATLAS; (ii) $100  \gev \le m_A \le 1000 \gev$, as a generic heavier pseudoscalar.
A pseudoscalar with $m_A \sim 95 \gev$ tends to give a contribution to the $pp \to Zh$ cross section that is too small
to set new limits on the parameter spaces of the two models. Heavier pseudoscalars, on the other hand, can yield
cross-section contributions larger than allowed by current LHC measurements and thus restricting
regions of parameter space of these models which are in agreement with theoretical and all other experimental 
constraints. The increase of LHC luminosity will yield even substantially tighter constraints. 
\end{abstract}

\begin{flushright}
\texttt{IFT-UAM/CSIC-26-031}
\end{flushright}

\newpage
\pagebreak


\section{Introduction} 
\label{sec:intro}

The last remaining elementary particle predicted by the Standard Model (SM), a Higgs boson,
was discovered by ATLAS and CMS in 2012 with a mass of $\sim 125 \gev$~\cite{ATLAS:2012yve,Chatrchyan:2012xdj}. 
In fact, detailed studies
of the interactions of this new particle have shown that the scalar discovered at the LHC is behaving very 
much as one would expect from the SM Higgs 
boson~\cite{CMS:2022dwd,ATLAS:2022vkf}. But such agreement with SM expectations
does not preclude the existence of a richer scalar sector -- current experimental uncertainties leave ample
room for deviations from SM-like behaviour of the 125 GeV scalar. Bounds from direct searches for extra scalars
have constrained the parameter space of models with extended scalar sectors, but by no means excluded them. In 
particular, the possible existence of an elementary pseudoscalar -- predicted in many beyond the SM (BSM) theories --
is an open question. Indeed, any theory that increases the number of $SU(2)$ doublets in the scalar sector
will necessarily predict the existence of at least one physical pseudoscalar. The simplest such example is the 
Two Higgs Doublet model (2HDM)~\cite{Lee:1973iz,Branco:2011iw}, 
which was originally proposed to have CP violation
as the result of spontaneous symmetry breaking. Doubling the number of scalar doublets yields (in a vacuum
which preserves CP) a pair of charged scalars and three neutral ones: two CP-even scalars $h$ and $H$ 
(one of which is expected to be similar to the SM Higgs) and a CP-odd state, the pseudoscalar $A$ 
(see~\cite{Branco:2011iw} for a review). 
The same is true  in Supersymmetric theories (such as the (N)MSSM)~\cite{Haber:1984rc,Gunion:1984yn,Ellis:1988er,Maniatis:2009re,Ellwanger:2009dp}, or models with
two doublets and an added scalar singlet, real or complex, 
e.g. the N2HDM~\cite{Grzadkowski:2009iz,Chen:2013jvg,Drozd:2014yla,Muhlleitner:2016mzt,Biekotter:2021ysx,Ferreira:2019iqb,Engeln:2020fld}
or the 2HDMS~\cite{Branco:2011iw,Baum:2018zhf,Heinemeyer:2021msz}, where the latter adds a second CP-odd 
Higgs boson to the spectrum.

There are several experimental anomalies that possibly also point towards an extended Higgs sector. One of the
most prominent examples are excesses around a mass value of $\sim 95 \gev$. The strongest excess was observed in the di-photon channel:
CMS has performed searches for (pseudo)scalar di-photon resonances below $125 \gev$, where the results based on the Run~1 and~2
results showed a local excess of $2.9\,\sigma$ at $95.4 \gev$~\cite{CMS-PAS-HIG-20-002}. Subsequently, ATLAS presented their full Run~2
result~\cite{ATLAS-CONF-2023-035} (focusing on their analysis with higher discriminating power) and found
an excess with a local significance of $1.7\,\sigma$ at precisely the same mass value, $95.4 \gev$. 
Combining these yielded a signal strength (neglecting possible correlations)~\cite{Biekotter:2023oen}, 
\begin{equation}
    \mu^\mathrm{exp}_{\gamma\gamma} = 0.24^{+0.09}_{-0.08}\,,
    \label{mugaga}
\end{equation}
corresponding to an excess with a local significance of $3.1\,\sigma$. 
More than 20 years ago, LEP reported a local $2.3\,\sigma$ excess in the~$e^+e^-\to Z(\phi\to b\bar{b})$ 
channel~\cite{LEPWorkingGroupforHiggsbosonsearches:2003ing}. The LEP result, taking into account the coarse mass resolution of 
the $b\bar b$ final state, is consistent with a Higgs boson with a mass of $95.4 \gev$ with a signal strength
of~\cite{Cao:2016uwt,Azatov:2012bz}
\begin{align}
    \mu_{b\bar{b}}^{\rm exp} = 0.117 \pm 0.057\,.
\label{mubb}
\end{align}
Finally, CMS observed another excess compatible with a mass of $95.4 \gev$ in the search for $pp \to \phi \to \tau^+\tau^-$~\cite{CMS:2022goy}. 
At $95 \gev$ this excess has a local significance of $\sim 2.4\,\sigma$ and a signal strength of
\begin{align}
    \mu^{\rm exp}_{\tau\tau} = 1.2 \pm 0.5~.
\label{mutautau}
\end{align}
ATLAS has not yet published a corresponding search. It should however be noted that the signal strength in \refeq{mutautau} 
is in tension with experimental bounds from recent searches performed by CMS for the production of a scalar Higgs boson 
in association with a top-quark pair or in association with a $Z$~boson, with subsequent decay into a tau pair~\cite{CMS-PAS-EXO-21-018}. 
Furthermore, corresponding searches at LEP for the process 
$e^+e^-\to Z(\phi\to\tau^+\tau^-)$~\cite{LEPWorkingGroupforHiggsbosonsearches:2003ing} do not yield support to the excess 
in the $\tau\tau$ channel.

\smallskip
There are several interesting phenomenological 
aspects about pseudoscalars which make their study interesting. For instance, production of such particles through
gluon fusion can be enhanced when compared to a CP-even scalar of the same mass, due to the presence of a 
$\gamma_5$ matrix in the pseudoscalar Yukawa coupling. 
This made it possible that a 95 GeV pseudoscalar from the 2HDM type~I could be used~\cite{Azevedo:2023zkg} 
to explain the di-photon and di-tau excesses described above.%
\footnote{
Analyses taking into account the combined CMS/ATLAS result in the di-photon channel (mostly together with the excess in the $b\bar b$ 
channel)
can be found in \citeres{Biekotter:2023oen,Belyaev:2023xnv,Aguilar-Saavedra:2023tql,Dutta:2023cig,Ellwanger:2023zjc,Cao:2023gkc,Arcadi:2023smv,Ahriche:2023wkj,Chen:2023bqr,Dev:2023kzu,Liu:2024cbr,
Cao:2024axg,Kalinowski:2024uxe,Ellwanger:2024txc,Ellwanger:2024vvs,Arhrib:2024zsw,Benbrik:2024ptw,Lian:2024smg,Ghosh:2025ebc,
Benbrik:2025hol,Chen:2025vtg} (see also \citeres{Biekotter:2017xmf,Domingo:2018uim,Biekotter:2019kde,Biekotter:2021qbc,Biekotter:2022jyr,
Biekotter:2023jld} for several early analyses of the excesses in the di-photon and $b \bar b$ channels).
}%
~Within the 2HDMS type~II it is demonstrated in \citere{2hdms-sh} that a singlet-like CP-odd Higgs boson can fit well the 
di-photon excesses.
On the other hand, the $b \bar b$ excess at LEP cannot be described by a pseudoscalar, as
the only possible production mechanism there would have been  
through a $ZZA$ vertex, but such a vertex is forbidden on CP symmetry grounds. 

Other interactions involving $A$ and $Z$ are allowed, though, and in this work we will focus
on the triple vertex $AZh$. This interaction emerges from the scalars' gauge kinetic terms, and allows for a 
pseudoscalar contribution to $Zh$ production, one of the mechanisms of Higgs production which have been probed
at the LHC  (as will be discussed below). In the SM the leading order contribution to this LHC production process is via quark-antiquark
annihilation into an off-shell $Z$ boson, $q\bar{q} \,\to\,Z\,\to\, Zh$, and therefore this process is suppressed
by the proton's PDFs, as the probability of finding antiquarks inside a proton is rather small. A pseudoscalar
raises an interesting possibility, as its production via gluon-gluon fusion is not PDF-suppressed and may be
quite sizeable; the leading order process is $gg\,\to\,A\,\to\,Zh$, which would have no interference
with the SM process and appears as a purely additive contribution to the SM-type $Zh$ production cross section. Thus a
sufficiently precise measurement of the cross section for $pp\,\to\,Zh$ can in principle be used to probe the contribution of a 
pseudoscalar and possibly put constraints on the parameter space of several models wherein such a particle is predicted.
This can be a pseudocalar at $95 \gev$, $A_{95}$ (see above), or more generaly a heavier CP-odd Higgs boson.
Both options will be explored in this work.

\smallskip
Concerning the experimental situation, 
at the LHC both ATLAS and CMS are measuring the $Zh$ production cross section. The latest ATLAS value using the
full Run~2 data is given by~\cite{ATLAS:2025qxq},
\beq
\sigma (pp \to Zh)^{\mathrm{ATLAS}}\,=\, 0.80 \pm 0.14\;{\rm pb}\,,
\label{si-exp}
\eeq
in perfect agreement with the SM theory prediction~\cite{ATLAS:2025qxq}. On the other hand, CMS only quotes a value relative 
to a SM reference cross section~\cite{CMS:2025jwz}. Furthermore, no ATLAS/CMS combined result for Run~2 data exists. 
Consequently, in this paper we will stick to the value given in \refeq{si-exp}. For the end of the HL-LHC it is expected 
that the experimental precision in this observable improves to $\pm 7\%$~\cite{ATL-PHYS-PUB-2025-012}.

In the following sections, we explore the contribution of a pseudoscalar Higgs to $Zh$ production channel via the $AZh$ coupling and the current limits and prospects for detection at the HL-LHC. 
First, in \refse{sec:mods}, we briefly review the two models we will be studying, the 2HDM and the 2HDMS. 
We detail our calculation of pseudoscalar contributions to $Zh$ production in \refse{sec:azh-calc}, and apply those 
results to the case of a 95 GeV scalar, both for a 2HDM type~I in \refse{sec:95TI} and within the 2HDMS type~II in~\ref{sec:A95S}.
We then generalize our cross section calculations for pseudoscalar masses between $100 \gev$ and $1000 \gev$ in \refse{sec:genA},
applying those results to ascertain parameter space cuts possible in the context of the 2HDM type~I and~II. Differential
distributions, and their importance 
for applying SM $Zh$ measurements to BSM-type pseudoscalar contributions, are discussed in
\refse{sec:diffdist}. We conclude in \refse{sec:conc}, and present cross section numerical
results for a wide range of pseudoscalar masses in the appendix.


\section{Model realizations}
\label{sec:mods}

We will analyze the pseudoscalar contribution to $Zh$ production via  gluon fusion in two UV-complete models that provide a 
well-defined calculational framework. These are the Two Higgs Doublet Model (2HDM)~\cite{Lee:1973iz,Branco:2011iw} and the corresponding
extension by another complex singlet (2HDMS)~\cite{Branco:2011iw,Baum:2018zhf,Heinemeyer:2021msz}.
We will give the relevant details of these models in the following subsections.
However, we would like to stress that these models should be regarded only as some of the potential realizations of pseudoscalar 
Higgs bosons, and the general results we will describe in the following sections can be applied to any 
model containing such a particle.
As will be explained further ahead, the full calculation for both models will simply require a few {\em coupling
modifiers}. To fix notation, the 2HDM Feynman rules of the vertices required for our results are given by 
(dropping eventual signs and factors of the imaginary unit~$i$)
\bea
A\,\overline{f}\,\gamma_5\,f & : & \xi^{f}_A\,\frac{m_f}{v}\,, \nonumber \\ 
A\,h\,Z_\mu & : & \frac{g}{2 \cos\theta_W} \,\xi^{Zh}_A\, (p_h \,-\,p_A)_\mu\,,
\label{eq:xidef}
\eea
where $A$ is the pseudoscalar, $f$ is a given fermion and $m_f$ its mass; 
constant and $\theta_W$ the weak mixing angle; 
and $p_x$ stands for the 4-momentum of particle $x$. The $\xi_A$ coupling modifiers will be simple functions of 
the diagonalization angles of scalar mass matrices. They have different expressions in each model considered, which we 
review in the following subsections. 


\subsection{The 2HDM}
\label{sec:2hdm}

The  most general 2HDM scalar potential has a total of 11 independent real parameters. In the Yukawa sector, it
predicts tree-level flavour changing neutral currents (FCNCs) mediated by scalars, which are known experimentally to be 
very small. In order to achieve natural suppression of those FCNCs, one frequently imposes a $Z_2$ symmetry, wherein
the Higgs doublets $\Phi_1$ and $\Phi_2$ transform as
\begin{align*}
    \Phi_1 \to \Phi_1 \,\,\,\,\, \Phi_2 \to - \Phi_2. 
\end{align*}
This also considerably simplifies the scalar potential, which becomes~\cite{Branco:2011iw}
\bea
  V_\mathrm{2HDM} &=& m^2_{11} \Phi_1^{\dagger}\Phi_1 + m^2_{22} \Phi_2^{\dagger}\Phi_2-(m^2_{12} \Phi_1^{\dagger}\Phi_2+\mathrm{h.c.})  \,
    + \, \frac{\lambda_1}{2} (\Phi_1^{\dagger}\Phi_1)^2  + \frac{\lambda_2}{2} (\Phi_2^{\dagger}\Phi_2)^2  \nonumber \\
  & &  + \lambda_3 (\Phi_1^{\dagger}\Phi_1) (\Phi_2^{\dagger}\Phi_2) + \lambda_4 (\Phi_1^{\dagger}\Phi_2) 
  (\Phi_2^{\dagger}\Phi_1) + \left[ \frac{\lambda_5}{2} (\Phi_1^{\dagger}\Phi_2)^2+\mathrm{h.c.}\right],
\eea
where all parameters are taken to be real. $m^2_{12}$ is a soft $Z_2$ breaking term, introduced so that the model 
may have, for convenience, a decoupling limit. Further, the doublets are parameterized as
\begin{equation}
\Phi_i = 
\begin{pmatrix}
  \phi^{\pm}_i  \\
  \frac{1}{\sqrt{2}}(v_i+ \rho_i + i \eta_i)\\
 \end{pmatrix}\,.
 \label{eq:doub}
\end{equation}
 The vacuum expectation values (\textit{vev}) of the Higgs doublets are denoted by $v_1$ and  $v_2$,
which obey $v =\sqrt{v^2_1+v^2_2} \simeq 246  \gev$ so that correct electroweak symmetry breaking occurs. 
 We  define $\tan \beta = v_2/v_1$
 so that one has $v_2 =v\sin\beta$ and $v_1 =v \cos \beta$. These \textit{vevs} are related to the parameters
 of the potential through the minimization equations, 
\begin{eqnarray}
 m^2_{11}= m^2_{12}\frac{v_2}{v_1} - \lambda_{1} \frac{v^2_1}{2} - \lambda_{345} \frac{v^2_2}{2},  \\
 m^2_{22}= m^2_{12}\frac{v_1}{v_2} - \lambda_{2} \frac{v^2_2}{2} - \lambda_{345} \frac{v^2_1}{2},
\end{eqnarray}
with
\begin{equation}
    \lambda_{345}=\lambda_3+\lambda_4+\lambda_5.
\end{equation}
  
 After electroweak symmetry breaking,  three Goldstone bosons are eaten up by the $W^{\pm}$ and $Z$ bosons while leaving two CP-even 
 (scalar) Higgs bosons $h$ and $H$, a CP-odd (pseudoscalar) Higgs boson  $A$ and a pair of charged Higgs mass eigenstates $H^{\pm}$ 
 in the CP conserving case.  
 There are several ways to extend the $Z_2$ symmetry to the fermion sector, yielding
 different ``types" of models.
For the 2HDM type~I, all quarks and leptons couple to the same Higgs doublet $\Phi_2$~\footnote{This is
a convention, we could have equally chosen $\Phi_1$ without any physical consequences.}.   
Thus, the Yukawa Lagrangian is 
\cite{Branco:2011iw} 
 \begin{eqnarray}
 \mathcal{L}^{I}_{\rm Yukawa}= -(y_u^{ij}Q_i \tilde{\Phi}_2 u_j+y_d^{ij}Q_i \Phi_2 d_j+y_l^{ij}L_i\Phi_2^{ij}l_j) + \text{ h.c.}, 
\end{eqnarray}  
where $i,j=1,2,3$ are the family indices of the fermions, $y_f$ 
($f=u,d,l$) are the Yukawa coupling matrices for the quarks and leptons and $\tilde{\Phi}_2 = i\sigma_2 \Phi^*_2$. For the 
2HDM type~II, the down-type quarks and leptons couple to $\Phi_1$ while the up-type quarks couple to $\Phi_2$ and the Yukawa 
Lagrangian is 
\begin{eqnarray}
 \mathcal{L}^{II}_{\rm Yukawa}= -( y_u^{ij}Q_i \tilde{\Phi}_2 u_j + y_d^{ij}Q_i \Phi_1 d_j + y_l^{ij}L_i\Phi_1^{ij}l_j)+ \text{ h.c.} 
\end{eqnarray}  
In the mass basis, the Yukawa Lagrangian can be written as~\cite{Branco:1996bq}
\begin{eqnarray}
   -\mathcal{L}_{\rm Yukawa} &=&  - \sum_{f=u,d,\ell} \frac{m_f}{v} (\xi_{h_i}^f \bar{f}f h_i- i \xi_A^f  \bar{f}\gamma_5f A) \nonumber \\
   &&- \left[\frac{\sqrt{2}V_{ud}}{v}\, \bar{u} \left( m_u \xi_A^u \text{P}_L+ m_d \xi_A^d \text{P}_R \right)d H^+
  +\frac{\sqrt2m_\ell\xi_A^\ell}{v}\, \bar{\nu}_L^{}\ell_R^{}H^+
 +\text{h.c.}\right]\,,
 \label{eq:Yuk}
\end{eqnarray}
where $m_f$ stands for fermion mass and $h_i$ (for $i=1,2$) are the CP even Higgs bosons. The 
Yukawa coupling modifiers
$\xi^f_{h_i}$ are defined as in \refta{tab:couplings} for the 2HDM type~I and type~II, where $\alpha$ is the diagonalization
angle of the $2\times 2$ CP-even scalar mass matrix.
\begin{table}[ht]
    \centering
    \begin{tabular}{|c|c|c|}
         \hline
   Coupling Modifier & type~I & type~II \\
   \hline
   $\xi^u_h$ & $\cos \alpha/ \sin \beta $ & $\cos \alpha/\sin \beta$  \\
  $\xi^d_h$ &$ \cos \alpha/ \cos \beta$ & -$\sin \alpha/ \cos \beta$ \\
 $\xi^{\ell}_h$&$\cos \alpha/ \cos \beta$ & -$\sin \alpha/ \cos \beta$   \\
     $\xi^u_H$ &  $\sin \alpha/ \sin \beta$  &$\sin \alpha/ \sin \beta$  \\
       $\xi^d_H$ & $\sin \alpha/ \sin \beta$ & $\cos \alpha/\cos \beta$  \\
    $\xi^{\ell}_H$& $\sin \alpha/ \sin \beta$ & $\cos \alpha/\cos \beta$ \\
    $\xi^u_A$ & $\cot   \beta $ & $\cot   \beta $  \\
    $\xi^u_A$ & $-\cot  \beta$  &$\tan \beta$ \\
   $\xi^d_A$ &$-\cot  \beta$ & $\tan \beta$\\
   \hline
    \end{tabular}
    \caption{Yukawa coupling modifiers in the 2HDM type~I and type~II~\cite{Branco:2011iw}.}
    \label{tab:couplings}
\end{table}
  
The couplings between CP-even scalars and gauge bosons pairs are scaled by $\sin (\beta-\alpha)$ for the SM-like 
Higgs $h$ and $\cos (\beta-\alpha)$ for the BSM Higgs $H$, compared to the SM Higgs-boson coupling. 
The CP-odd pseudoscalar does not couple to gauge boson pairs  
due to CP conservation. The pseudoscalar may however couple to a single gauge boson along with a Higgs boson or charged Higgs boson,
{\em i.e}, the vertices $AZh, AZH$ and $H^{\pm}AW^{\mp}$ are allowed at tree-level from CP considerations.   
The coupling factors $AZh$ and $AZH$ are proportional to $\cos(\beta-\alpha)$ and $\sin (\beta-\alpha)$ respectively. Thus,
for this model we  have $\xi^{Zh}_A = \cos(\beta-\alpha)$\. The other two relevant coupling modifiers are 
$\xi^t_A$ and $\xi^b_A$ from \refta{tab:couplings}
for the type~I or~II models.


\subsection{The 2HDMS}
\label{sec:2hdms}

The field content of the 2HDMS extends the 2HDM by one complex singlet $S$. In our 
model definition (see \citeres{Heinemeyer:2021msz,2hdms-sh} for details) the doublets and singlets are parameterized by,
\begin{gather}
        \Phi_1=\begin{pmatrix}
                \chi_1^+\\\phi_1
        \end{pmatrix}=\begin{pmatrix}
                \chi_1^+\\v_1+\displaystyle{\frac{\rho_1+i\eta_1}{\sqrt{2}}}
        \end{pmatrix}\,,\quad
        \Phi_2=\begin{pmatrix}
                \chi_2^+\\\phi_2
        \end{pmatrix}=\begin{pmatrix}
                \chi_2^+\\v_2+\displaystyle{\frac{\rho_2+i\eta_2}{\sqrt{2}}}
        \end{pmatrix}\notag\\
        S=v_S+\frac{\rho_S+i\eta_S}{\sqrt{2}}\,. 
\end{gather}
One should note the different convention for the {\em vevs} employed here, compared with the 2HDM 
in~\refeq{eq:doub}.
The model is subject to the same $Z_2$ symmetry considered in the previous section, 
as well as an additional $Z_3$ symmetry:
\begin{align}
Z_2 &~:~ \Phi_1 \to \Phi_1, \qquad \Phi_2 \to -\Phi_2, \qquad S\to S\,, \\[.3em]
Z_3 &~:~ 
        \begin{pmatrix} \Phi_1\\ \Phi_2\\ S \end{pmatrix} \to 
        \begin{pmatrix} 
        1&      &       \\      &       e^{i2\pi/3}&    \\      &       &       e^{-i2\pi/3}
           \end{pmatrix}\,
           \begin{pmatrix} \Phi_1\\ \Phi_2\\ S
        \end{pmatrix}\,.
\end{align}
This results in the Higgs potential
\begin{equation}
        \begin{split}
                V_{\rm 2HDMS}=&m_{11}^2(\Phi_1^\dagger\Phi_1)+m_{22}^2(\Phi_2^\dagger\Phi_2)+\frac{\lambda_1}{2}(\Phi_1^\dagger\Phi_1)^2+\frac{\lambda_2}{2}(\Phi_2^\dagger\Phi_2)^2+\lambda_3(\Phi_1^\dagger\Phi_1)(\Phi_2^\dagger\Phi_2)\\
                &+\lambda_4(\Phi_1^\dagger\Phi_2)(\Phi_2^\dagger\Phi_1)+m_S^2( S^\dagger S)+\lambda'_1( S^\dagger S)(\Phi_1^\dagger\Phi_1)+\lambda'_2( S^\dagger S)(\Phi_2^\dagger\Phi_2)\\
                &+\frac{\lambda''_3}{4}( S^\dagger S)^2+\Big(-m_{12}^2\Phi_1^\dagger\Phi_2+\frac{\mu_{S1}}{6} S^3+\mu_{12} S\Phi_1^\dagger\Phi_2+\text{h.c.}\Big)~.
        \end{split}
        \label{eq:2hdmspot}
\end{equation}
The tadpole/minimization equations allow us to trade $m_{11}^2$, $m_{22}^2$ and $m_{S}^2$ 
by the three {\em vevs}, resulting therefore in 12 independent free parameters.

The charged Higgs sector of the 2HDM and the 2HDMS have the same structure. 
However, the additional singlet changes the neutral sector, by mixing with both the CP-even and CP-odd components
of the doublets. 
This generates an additional scalar Higgs and an additional pseudo-scalar Higgs. In total we have 3 scalar Higgs bosons 
$h_1$, $h_2$, $h_3$, the charged Higgs boson $H^{\pm}$, as well as 2 pseudo-scalar Higgs bosons $a_1$, $a_2$.
We use the conventions $m_{h_1}<m_{h_2}<m_{h_3}$ and $m_{a_1}<m_{a_2}$.
One obtains the CP-even Higgs-boson mass eigenstates by diagonalizing a $3 \times 3$ mass matrix, $M_S^2$. 
Since $M_S^2$ is symmetric, the diagonalization matrix is orthogonal, given by 
\begin{equation}
        R=\begin{pmatrix}
                c_{\alpha_1}c_{\alpha_2}& s_{\alpha_1}c_{\alpha_2}& s_{\alpha_2}\\
                -s_{\alpha_1}c_{\alpha_3}-c_{\alpha_1}s_{\alpha_2}s_{\alpha_3}& c_{\alpha_1}c_{\alpha_3}-s_{\alpha_1}s_{\alpha_2}s_{\alpha_3}& c_{\alpha_2}s_{\alpha_3}\\
                s_{\alpha_1}s_{\alpha_3}-c_{\alpha_1}s_{\alpha_2}c_{\alpha_3}& -s_{\alpha_1}s_{\alpha_2}c_{\alpha_3}-c_{\alpha_1}{{s_{a}}}_3& c_{\alpha_2}c_{\alpha_3}
        \end{pmatrix}~,
        \label{eq:rot}
\end{equation}
where $\alpha_1$, $\alpha_2$ and $\alpha_3$ are the three mixing angles and for simplicity we adopted the notation
$s_x = \sin x$, $c_x = \cos x$. The mass basis and the interaction basis are related by,
\begin{equation}
        \begin{pmatrix}
                h_1\\h_2\\h_3
        \end{pmatrix} = R\begin{pmatrix}
                \rho_1\\ \rho_2\\ \rho_S
        \end{pmatrix},\quad \mathrm{diag}\{m^2_{h_1},m^2_{h_2},m^2_{h_3}\}=R^T {M}^2_{S} R~.
\end{equation}

In the CP-odd sector, taking into account the neutral Goldstone boson, one needs two mixing angles for the diagonalization. 
The first one is, as in the 2HDM, the angle $\beta$. Additionally we define the angle $\alpha_4$ for pseudo-scalar Higgses, and the mixing matrix for CP-odd sector can be expressed as:
\begin{gather}
        R^A=\begin{pmatrix}
                -s_\beta c_{\alpha_4}& c_\beta c_{\alpha_4}& s_{\alpha_4}\\
                s_\beta s_{\alpha_4}& -c_\beta s_{\alpha_4}& c_{\alpha_4}\\
                c_\beta& s_\beta& 0
        \end{pmatrix}
        \label{eq:rota}
\end{gather}
with
\begin{gather}    
        \begin{pmatrix}
                a_1\\a_2\\\xi
        \end{pmatrix}=R^A\begin{pmatrix}
                \eta_1\\ \eta_2\\ \eta_S
        \end{pmatrix},\qquad\mathrm{diag}\{m^2_{a_1},m^2_{a_2},0\}=(R^A)^T {M}^2_{P} R^A~,
\end{gather}
where $\xi$ denotes the Goldstone boson and ${M}^2_{P}$ is the un-rotated CP-odd scalar mass matrix.

These rotation matrices allow us to express the free parameters of the Lagrangian in terms of the mass of all
scalar particles and the mixing angles.
As a result we have the following set of input parameters, 
\begin{equation}
        \tan\beta,\quad\alpha_{1,2,3,4},\quad m_{h_1},\quad m_{h_2},\quad m_{h_3},\quad m_{a_1},\quad m_{a_2},\quad m_{H^\pm},\quad v_S\,.
        \label{eq:inpmass}
\end{equation}

The couplings of the Higgs bosons to fermions and gauge bosons are given as follows. The charged Higgs bosons couple exactly as in the 2HDM.
For the neutral Higgs bosons we define the coupling modifiers
as the ratio between the 2HDMS Higgs coupling and the corresponding 
SM-Higgs coupling:
\begin{equation}
        c_{h_i ff} = \frac{g_{h_i ff}}{g_{H_\text{SM}ff}}~.
\end{equation}
The coupling modifiers of the CP-even Higgs bosons to fermions for Yukawa models type~I and~II are summarized in 
\refta{tab:fermioncoup}.
\begin{table}[htb]
        \centering
\renewcommand{\arraystretch}{1.6}
        \begin{tabular}{l|c|c}
                \hline
                &type~I& type~II \\
                \hline
                $c_{h_i tt}$& $\frac{R_{i2}}{\sin\beta}$& $\frac{R_{i2}}{\sin\beta}$ \\
                $c_{h_i bb}$& $\frac{R_{i2}}{\sin\beta}$& $\frac{R_{i1}}{\cos\beta}$ \\
                $c_{h_i \tau\tau}$& $\frac{R_{i2}}{\sin\beta}$& $\frac{R_{i1}}{\cos\beta}$ \\
                \hline
        \end{tabular}
        \caption{Coupling modifiers between Higgs bosons and fermions for models type~I and~II.}
        \label{tab:fermioncoup}
\renewcommand{\arraystretch}{1.2}
\end{table}
The couplings of the CP-odd Higgs bosons, $a_{1,2}$ can be obtained from \refta{tab:fermioncoup} by the replacement $R \to R^A$. 
Furthermore, for the  Higgs to gauge-bosons  coupling modifiers we have
\begin{equation}
        c_{h_iVV}=c_{h_iZZ}=c_{h_iWW}=\cos\beta R_{i1}+\sin\beta R_{i2}\,, \quad c_{a_iVV} = 0\,.
        \label{eq:bosoncoup}
\end{equation}

Following \citeres{Heinemeyer:2021msz,2hdms-sh}, we assume here the 2HDMS type~II, with $m_{a_1} = 95 \gev$.%
\footnote{
In \citeres{Heinemeyer:2021msz,2hdms-sh} a {\em scalar} Higgs boson at $\sim 95 \gev$ was analyzed, and it was argued that a fit to the 
excesses cannot be realized in the 2HDMS type~I. While for a pseudoscalar Higgs, 
in principle,  a 2HDMS type~I realization also appears to be possible, we stick to the type~II analyses as provided in the literature.
}
For this model realization one finds,
\begin{equation}
\xi^{t}_A\,=\,\frac{c_{\alpha_4}}{t_\beta}\;\;\; , \;\;\; 
\xi^{b}_A\,=\,-\,s_{\alpha_4}\,t_\beta\,.
\end{equation}
A careful manipulation of the gauge terms in the covariant derivatives
of the scalars yields
\beq
\xi^{Zh}_A\,=\,s_{\alpha_4}\,\left( c_{\beta - \alpha_1} c_{\alpha_3} \,+\, 
s_{\beta - \alpha_1} s_{\alpha_2} s_{\alpha_3} \right)\,.
\eeq
The case of vanishing mixing between singlets and doublets corresponds to 
$\alpha_1 \equiv \alpha$, $\alpha_2=\alpha_3 = 0$ and $\alpha_4 = \pi/2$, 
and yields $\xi^{Zh}_A = c_{\beta - \alpha}$, i.e.\ the 2HDM value 
for this coupling modifier.


\section{Pseudoscalar contributions to \boldmath{$Zh$} production}
\label{sec:azh-calc}

As a starting point, we
calculated the contribution to the process $gg \to A \to Zh$ (with $h$ being the Higgs discovered at the LHC) within the
2HDM type~I for%
\footnote{
For concreteness, in all 2HDM calculations here we set $m_{12}^2 = -2000 \gev^2$. This choice has no impact on the cross-section
values. See, however, the discussion in \refse{sec:genA1}.
}
$\tb = 1$ and $\cba = 0.1$, for $\sqrt{s} = 14 \tev$ at NNLO QCD. The LO QCD result was calculated
with Madgraph (version \texttt{MG5}$\mathtt{\_}$\texttt{aMC}$\_$\texttt{v3.5.7}~\cite{Alwall:2014hca,Frederix:2018nkq}) 
using the default parton distribution functions \texttt{NNPDF2.3}\cite{Ball:2013hta} and using the UFO model files 
for the 2HDM type~I~\cite{wu_yongcheng_2023_8207058}. 
The NNLO QCD cross-sections are then obtained by normalizing the LO QCD cross-section computed with Madgraph to the NNLO cross-section 
computed using SusHI~\cite{Harlander:2015xur,Harlander:2016hcx} using a $K$-factor (the ratio of the gluon fusion production 
cross-section of the pseudoscalar $A$ at NNLO -- here in the heavy top limit -- 
and the cross-section computed using at LO QCD), where one finds $K \simeq 2$.

Concretely, we calculated the individual contributions from the top ($\sigma_t$) and bottom ($\sigma_b$)
quarks, and the total cross section ($\sigma_{\mathrm{tot}}$), 
as shown in~\refta{tab:A95cs} for the example of $\mA = 95 \gev$.
\begin{table}[ht!]
\centering
\begin{tabular}{|c|c|c|}
\hline
$\displaystyle\frac{}{} \sigma_t$ (pb) & $\sigma_b$ (pb) & $\sigma_{\mathrm{tot}}$ (pb)  
\\
\hline
$\displaystyle\ 5.29\times 10^{-3}$ & 
$3.041\times 10^{-7}$ &
$5.304\times 10^{-3}$
 \\
\hline
\end{tabular}
\caption{\em NNLO QCD cross sections for $gg \to A \to Zh$, at $\sqrt{s}=14 \tev$, computed with Madgraph for the
2HDM type~I, with $\mA= 95 \gev$, $\tb = 1$, $\cba = 0.1$ and $m_H = m_{H^{\pm}}=500 \gev$. Top-only ($\sigma_t$), bottom-only ($\sigma_b$) and total ($\sigma_{\mathrm{tot}}$) cross sections (see text).}
\label{tab:A95cs}
\end{table}
These results, though obtained in the 2HDM type~I, can be trivially used to investigate contributions from a pseudoscalar 
$A$ to LHC $Zh$ production in other models, by rescaling each contribution above by the appropriate $A$ coupling modifiers.

For a generic model, the $gg\to A\to Zh$ cross section can then be computed from the 
results of \refta{tab:A95cs} 
as,
\beq
\sigma(gg\to A\to Zh)\,=\,100\,\left(\sigma_t\,{\xi^t_A}^2\,+\,\sigma_b\,{\xi^b_A}^2\,+\,\sigma_i\,{\xi^t_A} {\xi^t_A}\right)\,
{\xi^{Zh}_A}^2,
\label{eq:gencross}
\eeq
with the ``interference cross section" given by $\sigma_i = \sigma_t + \sigma_b - \sigma_{\mathrm{tot}}$.~\footnote{This choice 
of signs
is related to the negative sign of $\xi_b$ in the 2HDM type-I (see \refta{tab:couplings}) and the fact that the numbers 
of~\refta{tab:A95cs} were obtained for exactly that model.} 
The coupling modifiers $\xi^t_A$, $\xi^b_A$ and $\xi^{Zh}_A$ are defined in the previous sections.
The factor of 100 stems from the fact that the numbers of \refta{tab:A95cs} were
obtained for $\cba = 0.1$, and the cross sections scales with $\cos^2(\beta-\alpha)$. It should be noticed that this contribution
to $Zh$ production is supposed to add directly to the SM one, since there is no interference -- though both processes share a final state, 
the initial one is different in both cases, since in the SM $Zh$ production is, at tree-level, initiated by $q\bar{q}$ annihilation, 
whereas  the pseudoscalar process uses gluon-gluon fusion.%
\footnote{There is also a contribution in the SM from $gg \to Zh$, which is, however 
an order of magnitude smaller than the dominant $q\bar q \to Zh$ cross 
section~\cite{PhysRevD.38.1008,PhysRevD.42.2253,LHCHiggsCrossSectionWorkingGroup:2016ypw,LHCCernYellow} 
(see also the discussion in \refse{sec:diffdist}).} 


\section{Analysis for the \boldmath{\anf}}
\label{sec:a95}

In this section we analyze possible limits on a light pseudoscalar, the \anf, as motivated in \refse{sec:intro}. 
We investigate two models that accomodate a light pseudoscalar to
fit the observed $\gamma\gamma$ excesses.
The two models are the 2HDM type~I, where the data is taken from \citere{Azevedo:2023zkg} and the 2HDMS type~II with the data obtained 
from \citeres{Biekotter:2023oen,Heinemeyer:2021msz}.


\subsection{The 2HDM type~I}
\label{sec:95TI}

\begin{figure}[ht!]
  \centering
\includegraphics[height=7cm,angle=0]{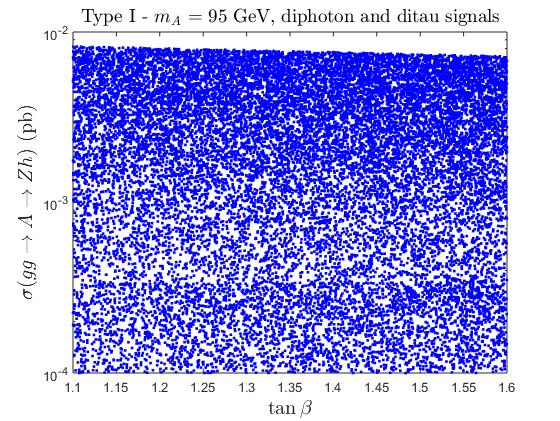}
  \caption{Contribution to $Zh$ production via gluon fusion to a pseudoscalar at LHC, for $m_A$ = 95 GeV, as a function 
  of $\tb$. The cross section is computed at NNLO.}
\label{fig:TypeI95}
\end{figure}

As a first concrete example,
we will consider the type-I interpretation of the di-photon excess at 95~GeV from~\citere{Azevedo:2023zkg}. In that 
paper it was shown that for $1.1 \lesssim \tb \lesssim 1.6$ it is possible to fit both the 95 GeV di-photon
excess observed at CMS and ATLAS, as well as the suggestion of a di-tau excess occurring for a similar mass value. 
Higgs precision data was considered as well, with $0.87\leq\kappa_f \leq 1.02$ and $0.98 \leq \kappa_V \leq 1$, with
the $\kappa$'s being the Higgs coupling modifiers from Tab.~\ref{tab:couplings}, which may also be expressed as 
\beq
\kappa_t\,=\,\kappa_b\,=\,\kappa_l\,=\,\sin(\beta - \alpha)\,+\,\frac{\cos(\beta - \alpha)}{\tb}\;\;\; ,\;\;\;
\kappa_V \,=\,\sin(\beta - \alpha)\,,
\label{eq:kappasI}
\eeq
for top, bottom and leptonic as well as gauge coupling modifiers. Scanning over this parameter space, considering the pseudoscalar
coupling modifiers in type~I given in \refta{tab:couplings} and computing $\sigma(gg\to A\to Zh)$
using \refeq{eq:gencross}, at NNLO QCD, 
we obtain the results shown in \reffi{fig:TypeI95},
%
%
We observe a maximum value of $\simeq$\,0.008\,pb, thus well below the current LHC precision for $Zh$ production which
at 1\,$\sigma$ is 0.14\,pb, see \refeq{si-exp}. 
The reason, clearly, is that $Zh$ production in this case, with $\mA = 95 \gev$, 
requires the intermediate pseudoscalar to be highly off-shell, and that penalizes the cross section. 
Thus the type-I interpretation of the di-photon excess at 95~GeV is currently not at all constrained by existing 
LHC data on $Zh$ production. Indeed, the pseudoscalar contribution for that process is negligible compared with the present experimental
uncertainty, we would need to achieve a precision of at least 1 \% on the $Zh$ production cross section at LHC to 
be sensible to
the presence of the \anf\ in that channel. Such accuracy is not expected even at the HL-LHC, where a precision 
of $\sim 7\%$ is anticipated~\cite{ATL-PHYS-PUB-2025-012} (Sections 4.1 and Fig 6 and 8 (right)).


\subsection{The 2HDMS type~II}
\label{sec:A95S}

In a 2HDM with an added complex singlet, a parameter scan of the type~II version of the
model was performed in \citere{2hdms-sh}. Compliance with all theoretical and experimental constraints was required, as
well as the presence of a pseudoscalar $a_1$ with mass equal to 95~GeV that could fit the di-photon excess
around that mass value. It was found that the 2HDMS type~II can easily describe the low-mass di-photon excesses.
The ``good'' data points found in \citere{2hdms-sh} were then used to perform the calculation
 %
\begin{figure}[ht!]
  \centering
\includegraphics[height=7cm,angle=0]{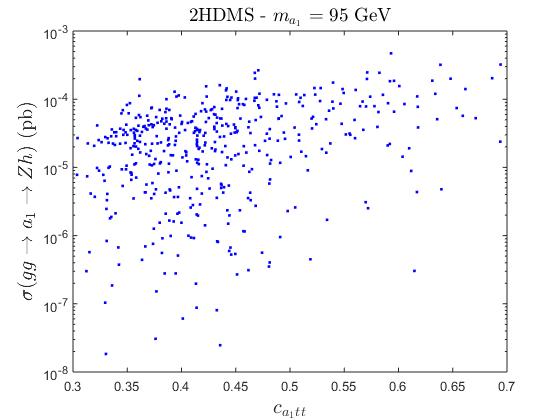}
  \caption{Contribution to $Zh$ production via gluon fusion to a pseudoscalar at LHC within the 2HDMS type~II,
  for $m_{a_1} = 95 \gev$, as a function 
  of the lightest pseudoscalar top coupling modifer $c_{a_1 tt}$ 
  The cross section is computed at NNLO (see text).}
\label{fig:2HDMS95}
\end{figure}
%
for $pp\to gg\to a_1 \to Zh$. The results are shown in \reffi{fig:2HDMS95} as a function of the pseudoscalar's top coupling modifier, 
$c_{a_1 tt}$ - this corresponds to the coefficient $\xi^t_A$, defined in \refeq{eq:xidef}. We observe that 
the predicted pseudoscalar contributions to $Zh$ production are even smaller than the 2HDM case, by themselves
already small. This is due to the fact that the pseudoscalar $a_1$ decaying to $Zh$ now has a singlet component, thereby 
reducing the $ha_1Z$ coupling w.r.t.\ the pure 2HDM case.


 \section{Constraints for generic \boldmath{$m_A$}}
 \label{sec:genA}

In the previous section we analyzed contributions to $Zh$ production via a pseudoscalar with a mass of 95~GeV, inspired by the
interpretation of possible di-photon and di-tau excesses hinted at by the LHC for that mass. 
While for this case the pseudoscalar contribution to the cross section turns out to be too small to set new limits on
the underlying model parameters, in this
section we take a different route: we investigate the pseudoscalar contribution for generic values of $m_A$. 

\subsection{General evaluation}
\label{sec:genA1}

\begin{figure}[ht!]
  \centering
 \includegraphics[height=7cm,angle=0]{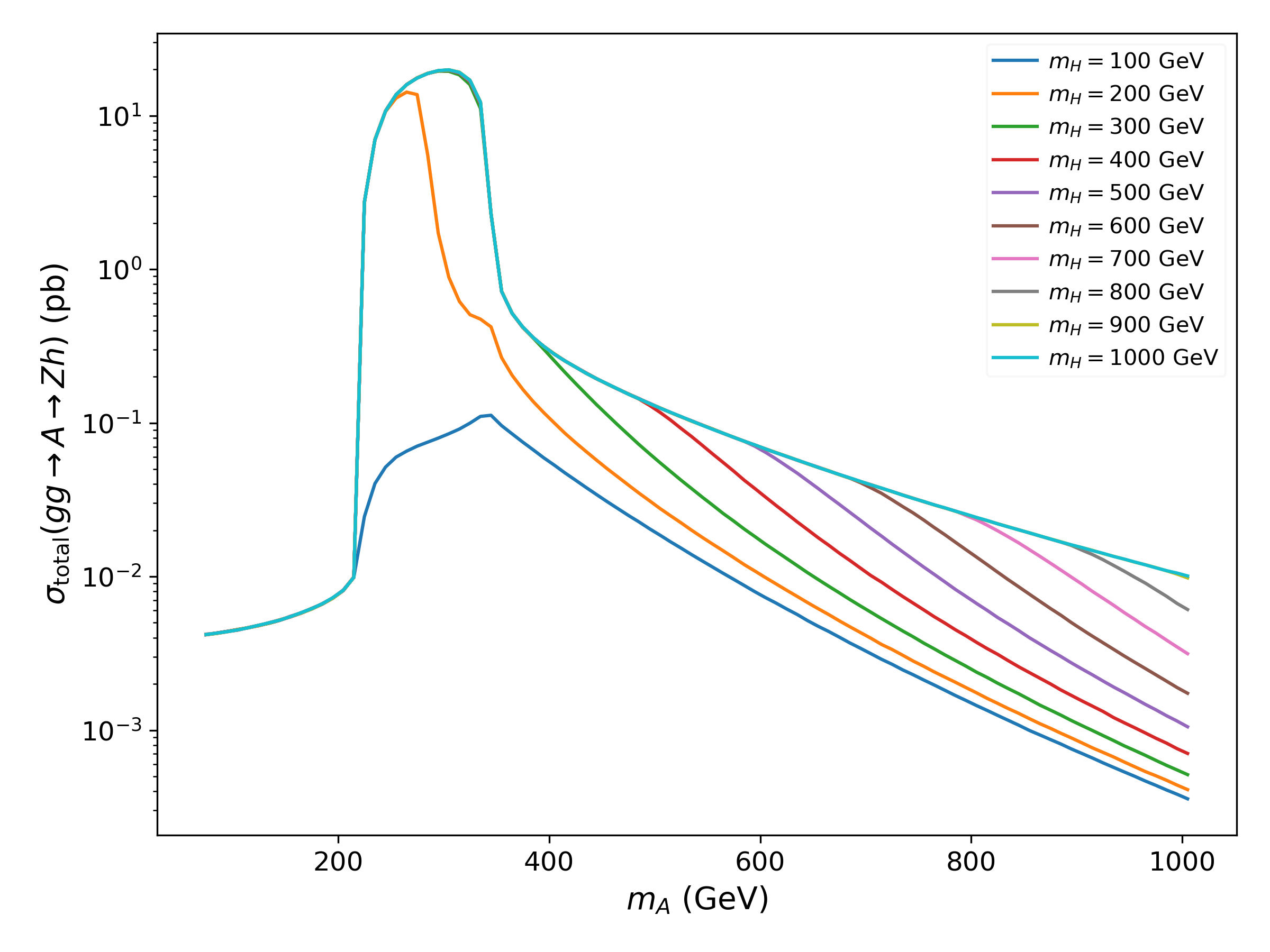}
  \caption{The cross section for $gg \to A \to Zh$ as a function of $m_A$ in the 2HDM type~I for $\sqrt{s} = 13$~TeV and
  for several values of $m_H = m_{H^\pm}$ (see color coding). The other parameters are $\tb = 1$, $\cba = 0.1$.}
\label{fig:cross_mA}
\end{figure}

We start with a general evaluation in
\reffi{fig:cross_mA}, where we show the cross section for the process
$gg\to A\to Zh$ as a function of $m_A$ in the 2HDM type~I for $\tb = 1$ and $\cba = 0.1$ at $\sqrt{s} = 13 \tev$.
The cross sections were obtained at LO QCD and multiplied by
%
%
a $K$ factor of $2$\, \textcolor{magenta}{(}see section \refse{sec:azh-calc}\textcolor{magenta}{)}.
The other parameters are chosen as follows: we set $m_H = m_{H^\pm}$ for several values in the range 100~GeV - 1000~GeV as 
given by the color coding. The choice of equal masses ensures
agreement with electroweak precision data~\cite{Chen:2019pkq}. We furthermore set $\tb = 1$ and $\cba = 0.1$ and $m^2_{12}=-2000$ GeV$^2$. 
For the purpose of this study the value of the soft breaking parameter is essentially
irrelevant -- it will have no direct impact on the couplings of the pseudoscalar $A$, and therefore, to good approximation, 
will not affect its production or decay. Indeed, by choosing specific values for the angles $\alpha$ and $\beta$ and
by fixing the masses of all scalars, $m^2_{12}$ then affects only the quartic couplings of the model. For instance, 
one finds that
\beq
\lambda_1\,=\,
\frac{1}{v^2 c_\beta^2}\left(c_\alpha^2 m_H^2 + s_\alpha^2 m_h^2
- m^2_{12}\frac{s_\beta}{c_\beta} \right) \,,
\eeq
and similar expressions for the other quartic couplings (see for instance~\cite{Branchina:2018qlf}). Such couplings are
crucial to compute the trilinear scalar couplings which determine, for instance, decays such as $H\to hh$, or the di-photon 
Higgs decay, $h\to \gamma\gamma$, but bear no direct contribution to the process $gg\to A \to Zh$.
The values of $\sigma_t$, $\sigma_b$ and $\sigma_{\mathrm{tot}}$ (following the notation employed in
\refta{tab:A95cs}) are presented in Appendix~\ref{ap:numbers} for $m_H = m_{H^\pm} = 1000 \gev$  and are available in
Ref.~\cite{dutta_2026_17558204} for the remaining $m_H$ values.

It is expected that the pseudoscalar $A$ will only give
significant contributions to $Zh$ production if $m_A \geq m_Z + m_h \simeq 216 \gev$  -- this is exactly
what is observed in \reffi{fig:cross_mA},
as the cross sections only achieve sizeable values above this threshold. However, if other decay channels 
of the $A$ open up, the 
respective branching ratios may well dominate over $A\to Zh$ and reduce the value of $\sigma(gg\to A\to Zh)$. 
In \reffi{fig:cross_mA} one can observe that 
as soon as the $t\bar{t}$ threshold is crossed, i.e.\ for $m_A \geq 2 m_t$, and the decay channel $A\to t\bar{t}$ becomes
allowed, the branching ratio for that channel will dominate (at least for small $\tb$) and the $Zh$ cross section drops significantly.
Likewise, if $m_H$ and $m_{H^\pm}$ are such that the decays $A\to ZH$ or $A\to W^\pm H^\mp$ are kinematically allowed, the respective
branching ratios are dominant: even with $A$ on-shell, the cross section for $gg\to A\to Zh$ is significantly reduced. 
This is exactly the behavior that can be observed in \reffi{fig:cross_mA}.
In summary, from this generic calculation one 
expects that the contribution from $A \to Zh$ production can set limits on the 2HDM parameter space if
$m_Z + m_h \leq m_A \leq 2 m_t$ and $\{m_H,m_{H^\pm}\} \gtrsim 300 \gev$.


\subsection{\boldmath{$A \to Zh$} constraints for a 2HDM type~I}

\begin{figure}[t]
\begin{tabular}{cc}
\includegraphics[height=6cm,angle=0]{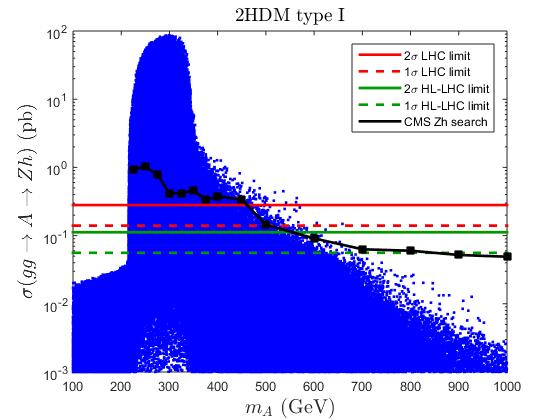}&
\includegraphics[height=6cm,angle=0]{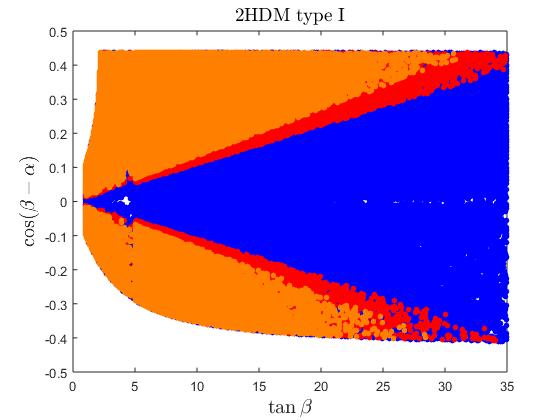}\\
 (a) & (b)
\end{tabular}
\caption{Scatter plots for the  parameter space of the 2HDM type~I (see text).
(a) NNLO Cross section for $gg\to A\to Zh$ as a function of $m_A$. The
solid/dashed red (green) line is the current lower  $2\,\sigma$/$1\,\sigma$ limit on $Zh$ production from LHC (HL-LHC).
The black line corresponds to the current limits of the  $gg\to A \to Zh$ cross section from CMS.
(b) $\tb$ vs.\  $\cos(\beta - \alpha)$: red points indicate those regions of parameter space above the $2\,\sigma$ 
solid red line, but below the CMS exclusion in the left plot; 
orange points indicate those regions of parameter space above the CMS exclusion limit in the left plot.
}
\label{fig:TI}
\end{figure}

We now repeat the cross section calculation as performed in \refse{sec:95TI}, 
but setting $100 \gev \leq m_A \leq 1000 \gev$, varying randomly in parallel $1\leq \tb\leq 50$
and $|\cba|\leq 0.5$. We also varied $300 \gev \lesssim \{m_H,m_{H^\pm}\} \lesssim 1000 \gev$  assuming $m_H=m_{H^{\pm}}$.
To comply with LHC Higgs precision constraints, we demanded (conservatively) constraints on
the Higgs coupling modifiers (shown in \refeq{eq:kappasI}) -- namely, all of them are required to lie in the interval 
$[0.9\,,\,1.1]$
(which directly yields $|\cba| \leq 0.44$). However, for this exploratory study on the $A \to Zh$ constraint we did not
test the compliance of our parameter points with the existing LHC bounds from heavy Higgs-boson searches, see the discussion in 
\refse{sec:diffdist}. 
For the cross-section calculation we used the results corresponding to $m_H = 1000 \gev$ to compute $\sigma(gg\to A\to Zh)$ and rescale the cross-sections to NNLO with a $K$-factor of 2.
The results are shown in \reffi{fig:TI}: in the left plot, showing the cross section predictions as a function of $m_A$,
one can observe, as expected, that there are only sizeable contributions from $A \to Zh$
%
%
production in the mass window discussed above. 
We also plot, in black, the direct observation limit on $gg\to A \to Zh$, see Fig.~4 in \citere{CMS:2025bvl}.
Using the $2(1)\,\sigma$ LHC current uncertainty on the $Zh$ cross section, 
indicated by the red horizontal solid (dashed) bar in \reffi{fig:TI}(a),
the parameter space excluded by the cross section measurement at $2\,\sigma$, but {\em not} the direct search
is shown as red points in the $\tb$-$\cba$ plane in \reffi{fig:TI}(b). 
The orange points in that plot correspond to the points excluded by the direct search, i.e.\ above the black line in \reffi{fig:TI}(a). 
It is interesting to observe that up to pseudocalar
masses of roughly 450/500 GeV, the current $2 \sigma$/$1 \sigma$ limits of the $pp\to Zh$ cross section are more restrictive
than the direct observation limit from~\cite{CMS:2025bvl}. The HL-LHC expected uncertainties on this cross section,
indicated by the green lines in the left plot of \reffi{fig:TI}(a),
are seen to be more restrictive than the current direct detection limits, up to masses of 600/700 GeV for the
$2 \sigma$/$1 \sigma$ uncertainties foreseen. However, their real exclusion power could only be determined from the comparison 
with the {\em future} direct search limits, for which no projection is available.

One can furthermore observe in \reffi{fig:TI}(b)
that parameter points with $\tb$ as high as 35 are excluded by the $A \to Zh$ bound.
For larger values of $\tb$
the pseudoscalar coupling modifier, see \refeq{eq:kappasI}, becomes too small due to its scaling like $1/\tb$ .
Very small values of $|\cba|$ are also not affected by this observable, which is to be expected since 
$\sigma(gg\to A\to Zh)\propto \cos^2(\beta - \alpha)$. 
The fact that no part of the parameter space in the $\tb$--$\cba$ plane is fully excluded by the $A \to Zh$ constraint
is a result of the marginalization over the Higgs-boson masses.


\subsection{\boldmath{$A \to Zh$} constraints for a 2HDM type~II}

We now repeat the analysis described in the previous subsection (same parameter ranges for $m_A$, $\tb$, $\cba$,
same constraints on the Higgs coupling modifiers), but now for the 2HDM type~II.
Higher values of $m_{H^\pm}$ have to be considered to comply with the
$b\to s \gamma$ constraints~\cite{Haller:2018nnx}. In the same way as in type~I, EWPO bounds require $m_H \sim m_{H^\pm}$.
The major difference in this case is the fact that the coupling modifiers to bottom-type quarks and charged leptons 
are now changed, 
since though the up-quarks couple only to $\Phi_2$, as in a type~I model, the other fermions couple to $\Phi_1$. 
As a consequence, we now have for the light Higgs and $A$ coupling modifiers
\bea
\kappa_t & =&  \sin(\beta - \alpha)\,+\,\frac{\cos(\beta - \alpha)}{\tb}\,, \nonumber \\
\kappa_b = \kappa_l & =&  \sin(\beta - \alpha)\,-\,\cos(\beta - \alpha)\,\tb\,, \nonumber \\
\xi^t_{A} & = & \frac{1}{\tb}\,, \nonumber \\
\xi^b_{A} = \xi^l_A &=& \tb\,,
\label{eq:kappasII}
\eea
with no changes for $\kappa_V$ and $\xi_{Zh}$. 
The effect of the changes in $\kappa_{b,l}$ will manifest itself in the allowed parameter points requiring 
$0.9 \le \kappa_{t,b,\tau} \leq 1.1$. The main effect of $\xi^b_A$ will be the 
impact on the interference terms between bottom and top contributions to $gg\to A$ (while $\xi^l_A$ does not enter into our analysis).

\begin{figure}[ht!]
\begin{tabular}{cc}
\includegraphics[height=6cm,angle=0]{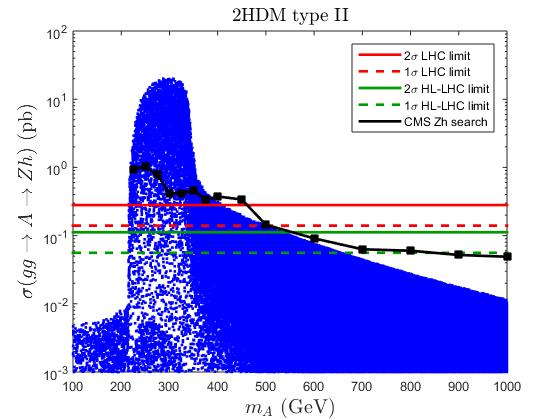}&
\includegraphics[height=6cm,angle=0]{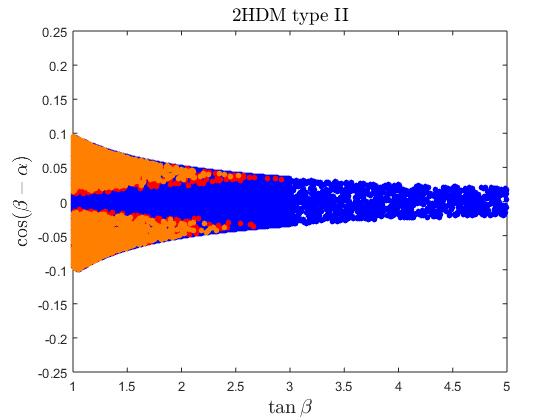}\\
 (a) & (b)
\end{tabular}
\caption{Scatter plots for the  parameter space of the 2HDM type~II (see text).
(a) NNLO Cross section for $gg\to A\to Zh$ as a function of $m_A$. The
solid/dashed red (green) line is the current lower  $2\,\sigma$/$1\,\sigma$ limit on $Zh$ production from LHC (HL-LHC).
The black line correspnds to the current limits of the  $gg\to A \to Zh$ cross section from CMS.
(b) $\tb$ vs.\  $\cos(\beta - \alpha)$: red points indicate those regions of parameter space above the $2\,\sigma$
solid red line, but below the CMS exclusion in the left plot; 
orange points indicate those regions of parameter space above the CMS exclusion limit in the left plot.
}
\label{fig:TII}
\end{figure}

The results are shown in \reffi{fig:TII}, as before in the $m_A$-$\sigma(gg \to A \to Zh$ plane (left) and in the
$\tb$-$\cba$ plane (right). The results in the left plot are very similar to the ones obtained for type~I. However,
the substantially different shape of the parameter space in \reffi{fig:TII}(b) as compared to \reffi{fig:TI}(b) reflects the
different form of $\kappa_{b,l}$ in both models, applied to the constraint $0.9 \le \kappa_{b,l} \le 1.1$.
The parameter space  thus allowed in the $\tb$-$\cba$ plane is now restricted by $A \to Zh$ only for very low values of $\tb \le 3$.
Other effects such as a change in the top- and bottom-loop interference in the $A$~production, or enhanced decay rates of 
$A \to bb, \tau\tau$ play only a minor role~\footnote{For clarity we truncate the plot in \reffi{fig:TII}(b)
at $\tb = 5$, though higher values of $\tb$ around $\cba = 0$ are allowed and were found in our scan.}.


\section{Differential Distributions}
\label{sec:diffdist}

\begin{figure}[ht!]
    \centering
     \includegraphics[scale=0.53]{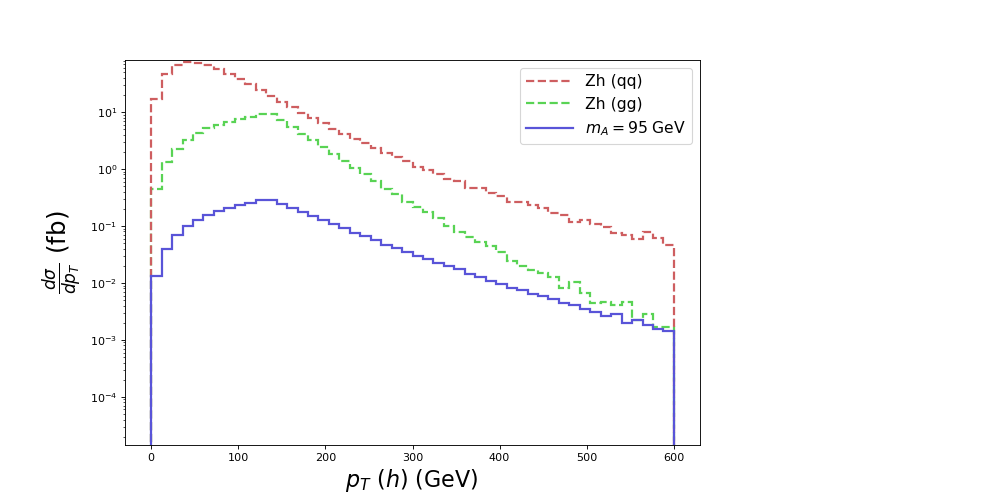} 
     \caption{Distributions of 
     the transverse momentum, $p_T$,      
     of $h$ (identical, at LO, to the $p_T$ of the $Z$) 
     from the signal and background (see text) for $m_A=95 \gev$
     for $\tb = 1.1$ and $\cba = 0.13$ at the parton level without any cuts.} 
    \label{fig:95GeVpT}
\end{figure}

\begin{figure}[ht!]
    \centering
    \includegraphics[scale=0.53]{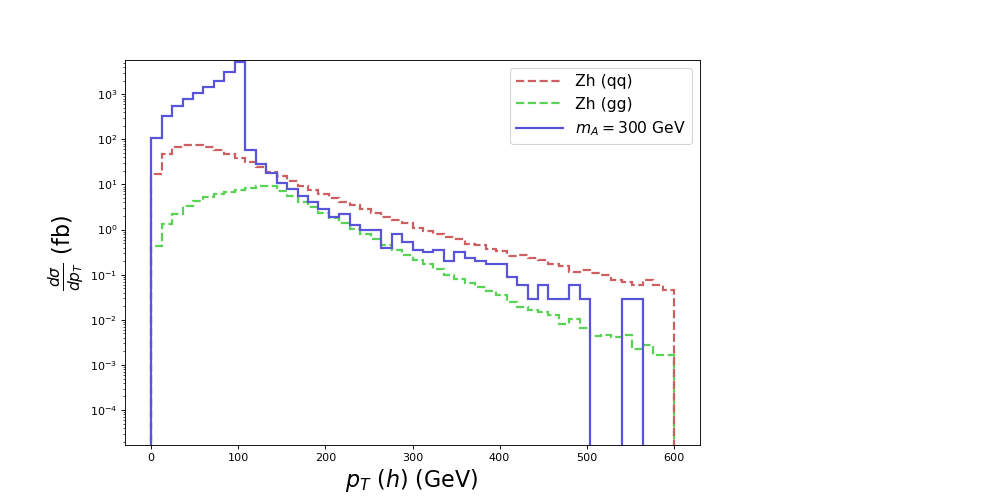}
     \caption{Distributions of $p_T$ of $h$ (identical, at LO, to the $p_T$ of the $Z$) 
    from the signal and background (see text) for $m_A = 300 \gev$, 
     $\tb = 1.1$ and $\cba = 0.4$ (above the red line in \protect\reffi{fig:TI}) at the parton level without any cuts. }  
    \label{fig:300GeVabovered}
\end{figure}

\begin{figure}[ht!]
    \centering
    \includegraphics[scale=0.53]{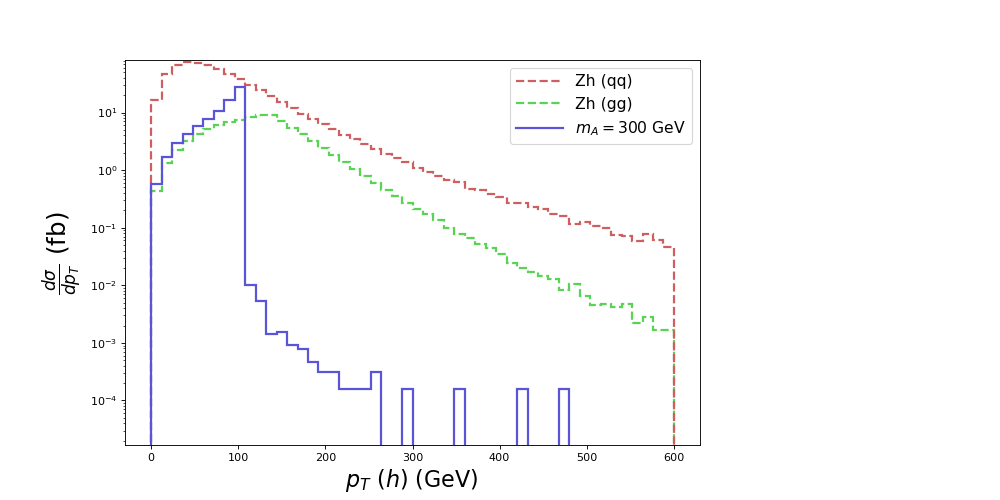}
    \caption{Distributions of $p_T$ of $h$ (identical, at LO, to the $p_T$ of the $Z$)  
    from the signal and background for $m_A=300 \gev$  
    for  
    $\tb = 1.1$ and $\cba = 0.005$ (below the red line in \protect\reffi{fig:TI}) at the parton level without any cuts. }  
    \label{fig:300GeVbelowred}
\end{figure}

In the previous sections we have applied experimental bounds on $Zh$ production to set limits on the allowed contribution
from $gg \to A \to Zh$. The application of these bounds, however, assumes that the differential distributions of the SM type
contributions, mainly from $q\bar{q} \to Z \to Zh$ (as used for the experimental analyses) 
are similar to the ones from $gg \to A \to Zh$. In this section we test that assumption and compare
the relevant differential distributions. In particular, we focus on
the differential cross-section distributions w.r.t.\ the transverse momentum ($p_T$) of the $Z$ and the Higgs boson $h$ at 
the partonic level at $\sqrt{s}=14 \tev$ for the SM background process $q\bar{q} \to Z \to Zh$ and the signal process $gg \to A \to Z h$. 
We will consistently quote the LO cross section as well as the one including available higher-order corrections, collectively 
denoted as NNLO.
\footnote{
We deviate from this setting, LO (NNLO) cross sections at $\sqrt{s} = 14 \tev$, only where we compare predictions 
to current measurements at the LHC, see \refta{tab:ptbins2} below. 
There we show the NNLO cross sections at $\sqrt{s} = 13 \tev$ including detector cuts.
}%
~The analysis is performed for benchmark points with $m_A = 95 \gev$ and $300 \gev$ in the 2HDM type~I. 
As for the color coding in our plots, for the SM background, the distributions arising from the process initiated by quark-antiquark
annihilation and that of gluon fusion to  $Zh$ production are shown in \reffis{fig:95GeVpT} - \ref{fig:300GeVbelowred}
as red and green dashed lines, respectively. 
The largest contribution to the SM $Zh$ process arises from the $q\bar{q}$ process, 
dominating over the $gg$ process by an order of 
magnitude~\cite{PhysRevD.38.1008,PhysRevD.42.2253,LHCHiggsCrossSectionWorkingGroup:2016ypw,LHCCernYellow}.
The signal distributions are shown as blue solid lines in all cases.
We only show the results for the $p_T$ of the Higgs, since at LO (i.e.\ without the radiation of another particle) the $2 \to 2$ 
process yields identical $p_T$ spectra for $Z$ and $h$. This changes somewhat once gluon radiation is taken into account, as was
shown in the SM in \citere{Gauld:2021ule,Davies:2026uxl}. 

We start in \reffi{fig:95GeVpT}, showing the distribution for a benchmark point with $m_A=95 \gev$ and the total 
 LO (NNLO) cross-section for $gg \to A \to Zh$ of 
3.63 (7.25)~fb, where the NNLO value is close to the maximum 
cross-section in \reffi{fig:TypeI95}.%
\footnote{The corresponding signal cross-section at $\sqrt{s}=13 \tev$ is 
3.06 (6.12)~fb, 
i.e.\ for $\sqrt{s} = 14 \tev$, an $\sim 18\%$ higher cross section is found.}
~While both the signal and background distributions follow a similar pattern, the cross sections differ by up to three orders 
of magnitude. Furthermore, while the SM background distribution has a peak around the $Z$-boson mass, 
the signal distribution has a peak around $\sim 150 \gev$ owing to the off-shell production of $Zh$ for $m_A=95 \gev$. 
Overall, it appears well justified to apply the limits obtained in the SM-like measurement of $pp \to Zh$ to the possible
$A_{95}$ contribution, albeit with the outcome that no limits on the $A_{95}$ can be obtained.

\begin{figure}[ht!]
     \centering
\includegraphics[scale=0.45]{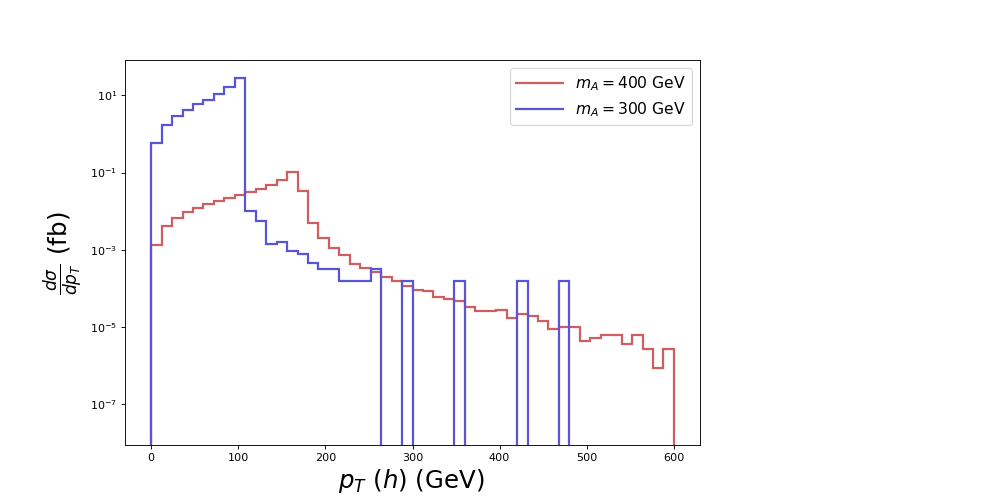}
     \caption{Distribution of $p_T$ of $h$ for two benchmarks with $m_A=300 \gev$ (blue) and $400 \gev$ (red) (at LO) for 
     $\tb = 1.1$ and $\cba = 0.005$. }
     \label{fig:pk}
 \end{figure}

\begin{figure}[ht!]
    \centering
     \includegraphics[scale=0.45]{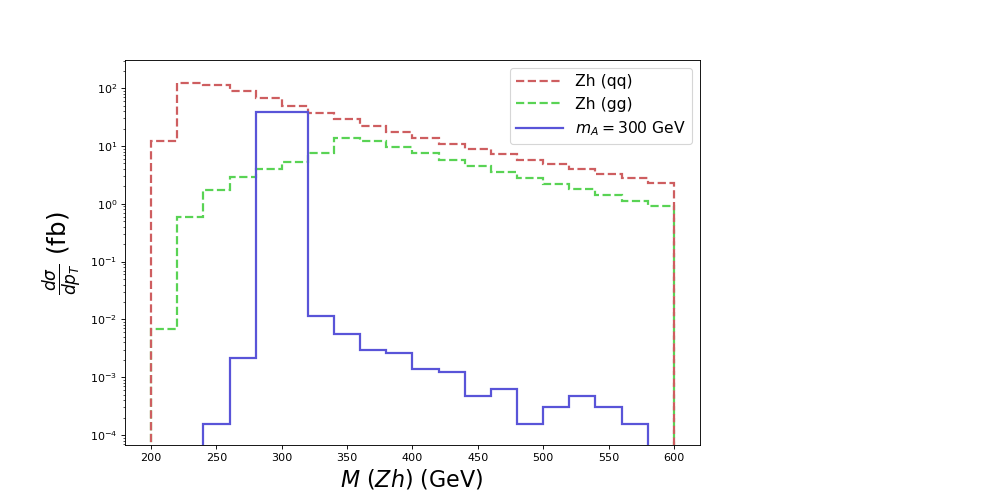}
    \caption{Distribution for the invariant mass of $Zh$ at parton level for signal and background  (at LO)
   without any cuts. The BSM point is our allowed benchmark point with $m_A = 300 \gev$.}
    \label{fig:mzh}
\end{figure}

The signal distribution changes significantly for a higher value of $m_A=300 \gev$. 
In \reffis{fig:300GeVabovered} and~\ref{fig:300GeVbelowred} we show two cases with $m_A=300 \gev$, 
both for $\tb = 1.1$, but with $\cba = 0.4$ and $0.005$, respectively. The two total cross sections for $gg \to A \to Zh$
are 14.7 (29.4)~pb at  LO (NNLO) and 0.0775 (0.155)~pb at LO (NNLO), respectively,
while for the SM, the total cross section is 0.732 (0.9861)~pb at LO (NNLO)%
\footnote{More precisely, the NNLO cross section contains NNLO QCD + NLO EWK for the Drell-Yan part $q\bar{q} \to Zh$ 
and $gg \to Zh + q\bar{q}$, as well as NLO + NLL QCD  for $gg \to Zh$) following \citere{LHCCernYellow}, which is in good agreement with Ref.~\cite{Davies:2026uxl}.
The corresponding signal cross-sections at $\sqrt{s}=13 \tev$ are 10.08 (20.16)~pb  and 0.067 (0.134)~pb at LO (NNLO), 
while for SM, the total $Zh$ cross-section is 0.659 (0.884)~pb at LO (NNLO).}%
. The higher (lower) cross section is located above (below) the current LHC limit shown as a red line in \reffi{fig:TI}(a), 
motivating the choice of these two benchmark points: the higher (lower) one corresponds to the maximum cross-section above (below) the red line,
i.e.\ the current LHC limits including the $2\,\sigma$ uncertainty, for $m_A=300 \gev$. The SM distributions, by definition, 
remain the same as those shown in our analysis of the $m_A = 95 \gev$ case. Concerning the $gg \to A \to Zh$ distributions,
in both cases, they follow a similar shape, however with a different normalization due to the reduced 
cross-sections for lower values of $\cba$. 
The signal distribution has a peak  and sharp edge around $100 \gev$. 
This is the Jacobian peak, which  is determined by the kinematics of the two body decay of $A\to Zh$, 
 i.e., via the K\"all\'en function,  beyond which the  differential cross-section falls. 
 We have checked for a higher mass of the pseudoscalar $A$, $m_A = 400 \gev$  (with all other parameters identical), that
 the Jacobian peak shifts to a larger value (due to the larger phase space available for the $A \to Zh$ decay), as can be
 seen in \reffi{fig:pk}.
 Naturally, no such sharp peak is observed in the SM background distributions for $q\bar{q}\to Zh$ around 100 GeV.
However, for the SM background, top-quark mass threshold effects lead to a broad peak in the $p_T$ distribution at $p_T\sim 150$ GeV for 
 $gg \to Zh$~\cite{PhysRevD.42.2253}.
We observe that for $p_T>200 \gev$  in all cases, the signal is lower than the background for all the benchmarks. On the other hand,
for lower $p_T$ bins, approximately below $130 \gev$,  the signal strongly exceeds the background in \reffi{fig:300GeVabovered} 
by several orders of magnitude, corresponding to our conclusion that this parameter point is already ruled out from current data.  
In \reffi{fig:300GeVbelowred} the signal cross section is smaller than the SM background for all bins, and it exhibits
a larger suppression for the higher $p_T$ bins. 
Finally, \reffi{fig:mzh} shows the invariant mass distribution of $Zh$, $m_{Zh}$, 
for the allowed benchmark with $m_A = 300 \gev$. As expected, for this benchmark, the dominant contribution to 
the production cross-section is from the resonance, and a peak appears at around $m_A = 300 \gev$.  
Conversely, it can be expected that with sufficient luminosity and a high enough resolution in $m_{Zh}$ the full cross section,
i.e.\ the SM together with the incoherent addition of the $gg \to A \to Zh$ contribution, a peak around $m_A$ should become visible, 
even if the total cross section does not differ in a significant way from the SM-only prediction.

\begin{table}[ht!]
    \centering
    \begin{tabular}{|l|c|c|c|c|}
     \hline
     $p_T$ [GeV]  & \multicolumn{4}{|c|}{Total $\sigma(Zh)$ at 14 TeV [fb] (LO)  for }  \\
     \cline{2-5} 
     \mbox{}\hspace{5mm}$m_A$ [GeV] &  95 &  300 (above) & 300 (below)   &  \multicolumn{1}{|c|}{SM}  \\
     \cline{1-5}   
    \hline 
       $<150$   & 2.07   &14657.09 &  77.69   & \multicolumn{1}{|c|}{634.36}  \\
      $150-250$    &1.11 & 31.3  &  3.89$\times10^{-3}$   & \multicolumn{1}{|c|}{81.28 }   \\
    $>250$  & 0.43  &5.61   &  0.93$\times10^{-3}$   & \multicolumn{1}{|c|}{18.86}     \\
    \hline  
    $p_T$ [GeV]  & \multicolumn{4}{|c|}{Total $\sigma(Zh)$ at 14 TeV [fb] (with $K$-factors)  for }   \\
     \cline{2-5}
       \hline 
       
       $<150$   & 4.15   & 29314.18  &  155.39   & \multicolumn{1}{|c|}{854.48}  \\
      $150-250$    &2.23  &  62.6  &  7.78$\times10^{-3}$   & \multicolumn{1}{|c|}{109.48 }   \\
    $>250$  & 0.86  & 11.22   &  1.86$\times10^{-3}$   & \multicolumn{1}{|c|}{25.40}     \\
    \hline  
      \end{tabular}
    \caption{STSX cross-section values in different $p_T(h)$ bins at the parton level at
   $\sqrt{s} = 14 \tev$  at LO (NNLO) in the upper (lower) part.}
    \label{tab:ptbins1}
\end{table}
%
\begin{table}[ht!]
\vspace{1em}
    \centering
    \begin{tabular}{|l|c|c|c|c|}
     \hline
       $p_T$ [GeV]  & \multicolumn{4}{|c|}{Total $\sigma(Zh)$ at 14 TeV [fb] (LO) ($|y_h|<2.5$)  for }  \\
     \cline{2-5} 
      \mbox{}\hspace{5mm}$m_A$ [GeV] &  95 &  300 (above) & 300 (below)   &  \multicolumn{1}{|c|}{SM}  \\
     \cline{1-5}   
       $<150$   &1.84  &13806.74 & 73.16  & \multicolumn{1}{|c|}{542.02}  \\
      $150-250$    &1.08& 30.65  &   3.58$\times10^{-3}$  & \multicolumn{1}{|c|}{76.93}   \\
    $>250$  &  0.42&   5.61&  0.93$\times10^{-3}$  & \multicolumn{1}{|c|}{18.52}     \\
       \hline
        $p_T$ [GeV]  & \multicolumn{4}{|c|}{Total $\sigma(Zh)$ at 14 TeV [fb] (with $K$-factors) ($|y_h|<2.5$)  for }  \\
     \cline{2-5} 
     \mbox{}\hspace{5mm}$m_A$ [GeV] &  95 &  300 (above) & 300 (below)   &  \multicolumn{1}{|c|}{SM}  \\
     \cline{1-5}   
    \hline 
       $<150$   & 3.86   &27613.48  & 146.33  & \multicolumn{1}{|c|}{730.11}  \\
      $150-250$    & 2.17 & 61.3  &   7.16$\times10^{-3}$  & \multicolumn{1}{|c|}{103.62}   \\
    $>250$  &  0.85&   11.22&  1.86$\times10^{-3}$  & \multicolumn{1}{|c|}{24.95}     \\
    \hline    
   
     \hline
    \end{tabular}
    \caption{STSX cross-section values in different $p_T(h)$ bins at the parton level at $\sqrt{s}=14 \tev$ with 
    a cut on the Higgs-boson rapidity of $|y_h| < 2.5$ at LO (NNLO) in the upper (lower) part.}
    \label{tab:ptbins1yh}
\end{table}

\begin{table}[ht!]
\vspace{1em}
    \centering
    \begin{tabular}{|l|c|c|c|c|c|}
     \hline
     $p_T$ [GeV]  & \multicolumn{4}{|c|}{
      $\sigma(Z (\to \ell\ell, \nu\nu)  h)$  at 13 TeV [fb] ($|y_h| < 2.5$) for }  &
     \multicolumn{1}{|c|}{Exp.\ measurements}\\
       \cline{2-5} 
     \mbox{}\hspace{5mm}$m_A$ [GeV] & 95  & 300 (above) & 300 (below)&   SM  & (CMS)~\cite{CMS-PAS-HIG-21-018}   \\
     \cline{1-6}   
     \hline
       $<150$   &1.02   & 6242.68  &39.31 & 205.36 & 200  \\
      $150-250$  & 0.570  &0.0028  & 0.003 & 28.37   &   32 \\
    $>250$  & 0.216  &  0.0024  & 0.00024   & 6.75 & 9.99   \\
     \hline
    \end{tabular}
    \caption{STSX cross-section values in different $p_T(h)$ bins at the parton level at NNLO (\i.e.\ 
    scaled with appropriate $K$-factors as discussed in the text)
    at $\sqrt{s}=13 \tev$ LHC with a cut on the Higgs-boson rapidity of $|y_h| < 2.5$
    and including the BR($Z \to \ell^+\ell^-, \nu_\ell \bar\nu_\ell$) ($\ell = e, \mu, \tau$).
The rightmost column shows the available measurements from CMS at LHC~\cite{CMS-PAS-HIG-21-018}. }
    \label{tab:ptbins2}
\end{table}

The overall question remains, whether the current (or future HL-LHC) limits can be readily applied to the $gg \to A \to Zh$ contribution.
Therefore, as a final step in our analysis we show the results for the differential distributions using the binning of 
Simplified Template Cross Sections (STXS)~\cite{Berger:2019wnu}, as used in the experimental analyses.
In \reftas{tab:ptbins1} and \ref{tab:ptbins1yh}
we present a rough estimate at the parton level of the production cross-section at $\sqrt{s}=14 \tev$ at the LHC 
without and with a rapidity cut $|y_h|<2.5$ (as employed in the experimental analysis)~\cite{CMS-PAS-EXO-21-018},
respectively.%
\footnote{To obtain the cross-section numbers to include higher orders  for the signal and background, we rescale the  numbers with the 
appropriate  $K$-factors as discussed in the text. The corresponding cross-section numbers for $\sqrt{s}=13 \tev$ are shown in \reftas{tab:pt13} 
and ~\ref{tab:pt13cuts}, respectively, in the appendix.}%
~It can be observed in both tables that in the case of $m_A = 95 \gev$ the BSM contributions in all three bins are far below the 
SM prediction, i.e.\ the the conclusion that these points are not affected by the current limits is confirmed.
Concerning the $m_A = 300 \gev$ (above) case, the first bin, $p_T < 150 \gev$, shows a huge value for the 2HDM as compared to the SM, 
confirming the exclusion of this point, while the other two bins show much less sensitivity. Finally, the $m_A = 300 \gev$ (below) case,
and focusing on the NNLO case, all three bins show 2HDM values substantially smaller than the SM numbers. 
In the first bin, $p_T < 150 \gev$, they reach up to $\sim 20\%$ of the SM values, i.e.\ roughly at the level of the future HL-LHC
uncertainty. Together with the shapes visible in \reffi{fig:300GeVbelowred}, we conclude that it appears justified to apply the 
existent bounds, based on a SM analysis, to the BSM case as well.

Finally, in \refta{tab:ptbins2} we present the 
$\sigma \times \mbox{BR}(Z)$ numbers at $\sqrt{s}=13 \tev$ for the different benchmarks considered above 
(folding in  BR($Z \to \ell^+\ell^-, \nu_\ell \bar\nu_\ell$) ($\ell = e, \mu, \tau$)) manually,
as well as applying an additional cut on the pseudo rapidity of the Higgs-boson, $|y_h| < 2.5$.
Alongside we compare these numbers to the current experimental measurements%
\footnote{No corresponding uncertainties are published.}
from CMS Run~2 data at LHC. 
CMS provides measurements of 
the $Zh$ cross-section (with $Z \to \ell^+ \ell^-, \nu_\ell \bar\nu_\ell$, ($\ell = e, \mu, \tau$)) 
in the different $p_T$ bins ranging from 0$-250 \gev$ 
and above, see Tab.~7 in \citere{CMS-PAS-HIG-21-018}.  
The corresponding results from ATLAS using Run~2 data are given in \citere{ATLAS:2022vkf},
which are in qualitative agreement with CMS.
The projected uncertainty from ATLAS and CMS combined with a luminosity of 3~ab$^{-1}$ in each experiment is 22\%~\cite{ATLAS:2025eii}.
For all three bins the SM numbers (including a $K$-factor)
are about at the same level as the CMS results, in agreement with the total cross section measurement being in good agreement with 
the SM prediction. 
Now, comparing to the three BSM pseudoscalar predictions,
one can observe that the cases $m_A=95 \gev$ and $300 \gev$ (below), 
are safe from the current limits. On the other hand, the point ``$m_A = 300 \gev$ (above)'', located above the red line 
in \reffi{fig:TI}, is excluded from the 0$-150 \gev$ bin data (as expected from \reftas{tab:ptbins1} and \ref{tab:ptbins1yh}, 
see the discussion above), confirming the application of the current bounds
to the $gg \to A \to Zh$ contribution. In all three cases the ``naive application'' of the total cross section bound agrees with 
the more detailed STXS analysis. 
Folding out the rapidity cut and the $Z$-boson decay, the BSM and SM numbers scale in the same way, i.e.\ any conclusion drawn from
the more detailed result can be taken over to the ``simpler'' result without cuts and BRs taken into account. However, we would like to
stress that a
detailed collider study will be required to further investigate the sensitivity 
in the allowed region particularly close to the red line in \reffi{fig:TI}.


\section{Conclusions}
\label{sec:conc}

The Higgs boson at $\sim 125 \gev$ has several possible ways of being produced at LHC, the most significant of which is gluon fusion. 
But associated production of a Higgs particle and a $Z$ boson is also possible, though with much smaller cross sections.
The measurement of this production cross section has been found to be in agreement with the value expected in the SM, albeit with a significant uncertainty. This channel is of particular 
interest for BSM theories that have an additional pseudoscalar in the spectrum:
that pseudoscalar $A$ can be produced via gluon fusion and then decay to $Zh$, potentially yielding a significant
contribution to the total cross section. Furthermore, since the LO production channel in the SM for $Zh$ is $q\bar{q} \to Zh$,
the pseudoscalar contribution, $gg\to A \to Zh$, will be additive to the SM one, since there can be no interference (at LO) 
between both processes. $Zh$ production could therefore be regarded as an observable sensitive to the presence of a pseudoscalar
in a model-independent manner, and yield bounds on BSM physics. 

In the first step of our analysis we investigated
possible deviations from SM expectations in $Zh$ production for a pseudoscalar
of mass of $\sim 95 \gev$, a scenario motivated by intriguing (apparent) excesses in di-photon (and possibly di-tau) searches
at the LHC. Our conclusion, though, is that for such low masses $Zh$ production requires the intermediate $A$ to be
strongly off-shell and the resulting cross sections are highly suppressed. We observed this behaviour for two different models,
the 2HDM type~I and the 2HDMS type~II, 
both of which have been used to explain the 95 GeV di-photon excess. Thus, if a pseudoscalar
is indeed a suitable explanation for the hitherto-unconfirmed $\sim 95 \gev$ excesses, this work predicts that it will cause
no relevant deviations in the $Zh$ cross section.

A thorough investigation of the contributions of pseudoscalar in the mass range of $100 \gev$ - $1000 \gev$
to $Zh$ production was then undertaken, computing the contributions
to the production cross section involving only top quarks in the gluon fusion process; bottom quarks only; and both of them
together, in the framework of a 2HDM type~I. This framework allows for an easy generalization of our results for arbitrary 
models with a pseudoscalar, since the individual contributions to $Zh$ cross sections can be scaled with appropriate 
coupling modifiers. As an example we applied our results to the 2HDM type~I and type~II, and showed that a sizeable portion of 
the parameter space of these models can already be excluded by the current precision in the $Zh$ cross section, going beyond
the published direct search limits. In particular,
we showed that in the region $m_h + m_Z \leq m_A \leq 2 m_t$ the possibility of exclusion is the highest: the lower bound stems 
from the need to have an on-shell pseudoscalar, so that a significant contribution to the cross section is achieved; and the 
higher bound pertains to the fact that above the di-top threshold the decay $A\to t\bar{t}$ becomes kinematically allowed
and drastically reduces the branching ratio for $A\to Zh$.

Finally we investigated whether it is justified to apply limits steming from SM-type $Zh$ measurements to a possible 
pseudoscalar contribution.
The presence of a pseudoscalar in an $s$-channel contribution to $Zh$ production was shown to cause observable differences
in the transverse momentum distributions of this process. In particular, we showed how in specific benchmarks wherein
even if the contribution of the pseudoscalar to the total $pp\to Zh$ cross section were small, specific bins of the $p_T$
differential cross sections may yield concrete evidence of the existence of $A$. 
Using the STXS bins currently employed by ATLAS and CMS, we confirmed that the ``naive application'' of the current bounds
to the $gg \to A \to Zh$ contribution is justified and agrees with the more detailed STXS analysis.
Though our analysis of differential cross 
sections was made at the parton level and for specific benchmarks, we argue that these observables merit a very
close look from our experimental colleagues, as they may be sensitive, for specific $p_T$ bins, to BSM physics in
the form of pseudoscalars.

\smallskip
It is planned to include the new limits into the public tool \texttt{HiggsBounds}, as part of the \texttt{HiggsTools} 
package~\cite{Bechtle:2008jh,Bechtle:2011sb,Bechtle:2013wla,Bechtle:2015pma,Bechtle:2020pkv,Bahl:2022igd,HT-new}.


\subsection*{Acknowledgments:} 

We thank H.~Bahl, M.~Cepeda, J.M.~Langford and C.~Li for helpful discussion.
We thank D.~Schieber and A.~Tutus for providing the 2HDMS data points.
J.D.\ acknowledges  support from The Institute of Mathematical Sciences, India. J.D.\ also acknowledges support from the National Science and Technology Council, Taiwan, and National Taiwan University, Taiwan.
P.M.F.\ is supported by \textit{Funda\c c\~ao para a Ci\^encia e a Tecnologia} 
through contracts UIDB/00618/2020, UIDP/00618/2020, CERN/FIS-PAR/0025/2021 and 2024.03328.CERN.
The work of S.H.\ has received financial support from the
grant PID2019-110058GB-C21 funded by
MCIN/AEI/10.13039/501100011033 and by ``ERDF A way of making Europe'', and in part by by the grant IFT Centro de Excelencia Severo Ochoa CEX2020-001007-S funded by MCIN/AEI/10.13039/ 501100011033.
S.H.\ also acknowledges support from Grant PID2022-142545NB-C21 funded by MCIN/ AEI/10.13039/501100011033/ FEDER, UE.


\appendix
\section{Appendix}
\subsection{STXS numbers at \boldmath{$13 \tev$}}
\label{ap:numbers}

Here we show the $13 \tev$ numbers corresponding to \reftas{tab:ptbins1} and \ref{tab:ptbins1yh}.

\begin{table}[ht!]
    \centering
    \begin{tabular}{|l|c|c|c|c|}
     \hline
         $p_T$ [GeV]  & \multicolumn{4}{|c|}{Total $\sigma(Zh)$ at 13 TeV [fb] (LO)  for }   \\
       \cline{2-5} 
     \mbox{}\hspace{5mm}$m_A$ [GeV] &  95  & 300 (above) & 300 (below)&  \multicolumn{1}{|c|}{SM}   \\
     \cline{1-5}   

        $<150$   & 1.76 & 10075.65  &   66.89 & \multicolumn{1}{|c|}{570.24}  \\
      $150-250$    & 0.94 &   2.94 &  4.68$\times10^{-3}$    & \multicolumn{1}{|c|}{ 71.73}   \\
    $>250$  & 0.70  &   0.81 & 8.02$\times10^{-4}$    & \multicolumn{1}{|c|}{16.51 } \\
     \hline
        $p_T$ [GeV]  & \multicolumn{4}{|c|}{Total $\sigma(Zh)$ at 13 TeV [fb] (with $K$-factors)  for }   \\
          \cline{2-5} 
     \mbox{}\hspace{5mm}$m_A$ [GeV] &  95  & 300 (above) & 300 (below)&  \multicolumn{1}{|c|}{SM}   \\
     \hline
       $<150$   & 3.53   & 20151.3  &   133.79 & \multicolumn{1}{|c|}{ 764.70}  \\
      $150-250$    & 1.89 &   5.89 &  9.36$\times10^{-3}$    & \multicolumn{1}{|c|}{ 96.19}   \\
    $>250$  & 0.70  &   0.81 & 8.02$\times10^{-4}$    & \multicolumn{1}{|c|}{22.14 }     \\
    \hline 
    \end{tabular}
    \caption{STSX cross-section values in different $p_T(h)$ bins at the parton level at
   $\sqrt{s} = 13 \tev$  without 
    a cut on the Higgs-boson rapidity  at LO (NNLO) in the upper (lower) part.}
    \label{tab:pt13}
\end{table}

\begin{table}[ht!]
\vspace{1em}
    \centering
    \begin{tabular}{|l|c|c|c|c|}
     \hline
       $p_T$ [GeV]  & \multicolumn{4}{|c|}{Total $\sigma(Zh)$ at 13 TeV [fb] (LO) ($|y_h| < 2.5$) for }   \\
       \cline{2-5} 
     \mbox{}\hspace{5mm}$m_A$ [GeV] &  95  & 300 (above) & 300 (below)&  \multicolumn{1}{|c|}{SM}   \\
     \cline{1-5}   
     \hline
       $<150$   &1.65 &9556.97  & 63.46  &  \multicolumn{1}{|c|}{494.49}   \\
      $150-250$  & 0.91 &2.92&  4.54$\times10^{-3}$ &\multicolumn{1}{|c|}{68.32}    \\
    $>250$  & 0.69 &  0.81 & 8.02$\times10^{-4}$  & \multicolumn{1}{|c|}{16.27}    \\
    \hline 
    
        $p_T$ [GeV]  & \multicolumn{4}{|c|}{Total $\sigma(Zh)$ at 13 TeV [fb] (with $K$-factors) ($|y_h| < 2.5$) for }   \\
       \cline{2-5} 
     \mbox{}\hspace{5mm}$m_A$ [GeV] &  95  & 300 (above) & 300 (below)&  \multicolumn{1}{|c|}{SM}   \\
     \cline{1-5}   
     \hline
       $<150$   &3.31 & 19113.94  & 126.93   &  \multicolumn{1}{|c|}{663.11}   \\
      $150-250$  & 1.82 &5.84&  9.09$\times10^{-3}$ &\multicolumn{1}{|c|}{91.62}    \\
    $>250$  & 0.69 &  0.81 & 8.02$\times10^{-4}$  & \multicolumn{1}{|c|}{21.82}    \\
    \hline 
    \end{tabular}
    \caption{STSX cross-section values in different $p_T(h)$ bins at the parton level at
   $\sqrt{s} = 13 \tev$  with 
    a cut on the Higgs-boson rapidity of $|y_h| < 2.5$ at LO (NNLO) in the upper (lower) part.}
    \label{tab:pt13cuts}
\end{table}

\subsection{Cross section numbers}
\label{ap:numbers}

Finally, we present our results for the 2HDM contributions to $Zh$ production via gluon fusion to a pseudoscalar, for a wide range
of values of the mass $m_A$. The calculation was undertaken for 2HDM type~I, but these results can easily be applied to other models,
as explained in \refse{sec:azh-calc}. In the table below the top-only, bottom-only and total cross sections are shown for 
$\tb = 1$ and $\cba = 0.1$ and $m_H = m_{H^\pm} = 1000 \gev$ -- this latter choice was done to close decay channels such as
$A\to ZH$ and $A\to W^\pm H^\mp$.

\begin{table}[ht!]
\caption{\em LO QCD LHC cross sections for $gg\to A\to Zh$ computed with \texttt{MG5}$\_$\texttt{aMC}$\_$\texttt{v3.5.7} at  
$\sqrt{s}=13 \tev$ and $14 \tev$ for the 2HDM type-I,
with $m_H = m_{H^\pm} = 1000 \gev$ , $\tb = 1$, $\cba = 0.1$ and $m^2_{12}=-2000 \gev^2$. Top-only ($\sigma_t$), bottom-only 
($\sigma_b$) and total ($\sigma_{\mathrm{tot}}$) cross sections are listed and all values are in $pb$.}
\label{tab:A95}
\footnotesize
\begin{center}
\begin{tabular}{|c|c|c|c|c|c|c|}
\hline
$m_A$ (GeV) & \multicolumn{3}{|c|}{Cross-sections at $\sqrt{s}=13 \tev$} &  \multicolumn{3}{c|}{Cross-sections at $\sqrt{s}=14 \tev$} \\ 
\cline{2-7}
&  $\sigma_t$ (pb)& $\sigma_b$ (pb)& $\sigma_{\mathrm{tot}}$(pb)& $\sigma_t$ (pb)& $\sigma_b$ (pb)& $\sigma_{\mathrm{tot}}$(pb) \\
\hline
75 & $2.09\times10^{-3}$&$1.19\times10^{-7}$ &$2.09\times10^{-3}$ & $2.48\times10^{-3}$ &$1.40\times10^{-7}$& $2.48\times10^{-3}$  \\
85 &$2.14\times10^{-3}$ & $1.23\times10^{-7}$& $2.14\times10^{-3}$&$2.53\times10^{-3}$ &$1.44\times10^{-7}$&   $2.53\times10^{-3}$   \\
95  &$2.19\times10^{-3}$ &$1.28\times10^{-7}$ & $2.19\times10^{-3}$&$2.59\times10^{-3}$ &$1.49\times10^{-7}$ &   $2.59\times10^{-3}$   \\
105  &$2.24\times10^{-3}$ &$1.34\times10^{-7}$ & $2.25\times10^{-3}$&$2.64\times10^{-3}$ &  $1.56\times10^{-7}$&  $2.68\times10^{-3}$  \\
115 & $2.30\times10^{-3}$&$1.40\times10^{-7}$ & $2.32\times10^{-3}$&$2.74\times10^{-3}$ & $ 1.64\times10^{-7}$&   $2.75\times10^{-3}$   \\
125  & $2.39\times10^{-3}$&$1.49\times10^{-7}$ & $2.41\times10^{-3}$& $2.83\times10^{-3}$&$1.73\times10^{-7}$ &  $2.85\times10^{-3}$  \\
135  & $2.48\times10^{-3}$&$1.59\times10^{-7}$ & $2.50\times10^{-3}$&$2.94\times10^{-3}$ &$1.85\times10^{-7}$&  $2.95\times10^{-3}$   \\
145  &$2.59\times10^{-3}$ & $1.71\times10^{-7}$& $2.60\times10^{-3}$& $3.08\times10^{-3}$& $ 1.99\times10^{-7}$&  $3.07\times10^{-3}$  \\
155  &$2.73\times10^{-3}$ &$1.86\times10^{-7}$ & $2.74\times10^{-3}$& $3.23\times10^{-3}$& $2.17\times10^{-7}$& $3.24\times10^{-3}$ \\
165 &$2.87\times10^{-3}$ &$2.06\times10^{-7}$ & $2.90\times10^{-3}$ & $3.38\times10^{-3}$&$2.42\times10^{-7}$ &  $3.39\times10^{-3}$ \\
175  & $3.07\times10^{-3}$& $2.34\times10^{-7}$& $3.08\times10^{-3}$& $3.59\times10^{-3}$ &$2.73\times10^{-7}$&   $3.65\times10^{-3}$  \\
185 & $3.31\times10^{-3}$&$2.71\times10^{-7}$ & $3.32\times10^{-3}$ &$3.87\times10^{-3}$&$3.15\times10^{-7}$ & $3.91\times10^{-3}$   \\
195 &$3.60\times10^{-3}$ & $3.28\times10^{-7}$& $3.64\times10^{-3}$ &$4.23\times10^{-3}$ &$3.83\times10^{-7}$& $4.26\times10^{-3}$   \\
205 &$4.03\times10^{-3}$ &$4.32\times10^{-7}$ &$4.06\times10^{-3}$  &$4.73\times10^{-3}$ &$5.02\times10^{-7}$ & $4.77\times10^{-3}$ \\
215  &$4.90\times10^{-3}$ &$7.63\times10^{-7}$ & $4.94\times10^{-3}$&$5.72\times10^{-3}$ &$8.86\times10^{-7}$& $5.80\times10^{-3}$ \\
225  &$1.32$ &$8.62\times10^{-4}$ & $1.36$&1.52 & $ 9.90\times10^{-4} $& 1.57   \\
235 &3.37 &$1.84\times10^{-3}$ &3.48  &3.88& $2.13\times10^{-3}$& 4.01  \\
245 &5.17&$2.39\times10^{-3}$ & 5.35& 6.00& $2.77\times10^{-3} $&  6.18  \\
255 &6.62 &$2.58\times10^{-3}$ & 6.83 &7.67 & $2.97\times10^{-3}$ &  7.89   \\
265 &7.74 &$2.53\times10^{-3}$ & 7.95 &8.97&$ 2.92\times10^{-3} $ &  9.22   \\
275 & 8.61&$2.35\times10^{-3}$ & 8.78 &9.96 & $2.73\times10^{-3}$&  $10.22$   \\
285 &9.20 &$2.11\times10^{-3}$ & 9.41 &10.66 &$2.44\times10^{-3}$&  $10.94$  \\
295 &9.59 &$1.81\times10^{-3}$ &  9.82&$11.15$ &$2.12\times10^{-3}$&  $11.39$  \\
305 & 9.74&$1.51\times10^{-3}$ &9.93  &$11.31$&$ 1.76\times10^{-3}$ &   $11.55$ \\
315& 9.38& $1.18\times10^{-3}$ & 9.56 & 10.96& $1.39\times10^{-3}$&  $11.17$  \\
325 &8.39 & $8.29\times10^{-4}$  &8.51 &9.79 & $9.67\times10^{-4}$&  9.93   \\
335 &6.04 &  $4.41\times10^{-4}$ & 6.13 &7.08 &$5.17\times10^{-4}$ & 7.16 \\
345 & 1.13 & $4.77\times10^{-4}$  & 1.14 & 1.33& $5.56\times10^{-5}$& $1.34$ \\
355 &3.57$\times10^{-1}$ &  $1.43\times10^{-5}$ & 3.59 $\times10^{-1}$ &$4.17\times10^{-1}$ &$1.67\times10^{-5}$ & $4.21\times10^{-1}$  \\
365 &2.57$\times10^{-1}$ & $9.97.\times10^{-6}$  & 2.58$\times10^{-1}$&$3.00\times10^{-1}$ & $1.17\times10^{-5}$ & $3.02\times10^{-1}$ \\
375 &2.09$\times10^{-1}$ & $7.78\times10^{-6}$  & 2.11$\times10^{-1}$  & $2.46\times10^{-1}$&$9.28\times10^{-6}$ & $2.47\times10^{-1}$  \\
385 & 1.78$\times10^{-1}$&  $6.59\times10^{-6}$ & 1.79 $\times10^{-1}$&$2.09\times10^{-1}$&$7.75\times10^{-6}$ &  $2.10\times10^{-1}$  \\
395 & 1.57$\times10^{-1}$& $5.67\times10^{-6}$  & 1.57 $\times10^{-1}$&$1.84\times10^{-1}$&$6.67\times10^{-6}$& $1.85\times10^{-1}$    \\
405& 1.40$\times10^{-1}$& $4.96\times10^{-6}$  & 1.40 $\times10^{-1}$ & $1.65\times10^{-1}$&$5.84\times10^{-6}$&  $1.65\times10^{-1}$   \\
415&1.26$\times10^{-1}$ & $4.39\times10^{-6}$  & 1.26 $\times10^{-1}$ &$1.49\times10^{-1}$& $ 5.18\times10^{-6}$& $1.49\times10^{-1}$  \\
425 &1.15$\times10^{-1}$ &  $3.92\times10^{-6}$ &1.15$\times10^{-1}$  &$1.36\times10^{-1}$ & $4.62\times10^{-6} $& $1.36\times10^{-1}$  \\
435 &1.05$\times10^{-1}$ & $3.52\times10^{-6}$  &1.05  $\times10^{-1}$&$1.25\times10^{-1}$&$4.15\times10^{-6}$ & $1.24\times10^{-1}$  \\
445 &9.72$\times10^{-2}$ &  $3.19\times10^{-6}$ & 9.69$\times10^{-2}$ &$1.15\times10^{-1} $ & $3.76\times10^{-6}$& $1.14\times10^{-1}$ \\
455 &8.94$\times10^{-2}$ & $2.90\times10^{-6}$  &  8.98$\times10^{-2}$&$1.06\times10^{-1}$&$3.43\times10^{-6}$&  $1.06\times10^{-1}$ \\
465 &8.29$\times10^{-2}$ & $2.65\times10^{-6}$  & 8.32$\times10^{-2}$ & $9.81\times10^{-2}$& $3.13\times10^{-6} $&   $9.82\times10^{-2}$   \\
475 &7.73$\times10^{-2}$ &  $2.42\times10^{-6}$ &7.69$\times10^{-2}$ & $9.15\times10^{-2}$&$2.87\times10^{-6}$&  $9.15\times10^{-2}$    \\
485 & 7.20$\times10^{-2}$&  $2.21\times10^{-6}$ & 7.18$\times10^{-2}$ & $8.55\times10^{-2}$& $2.62\times10^{-6}$& $8.52\times10^{-2}$  \\
\hline 
\end{tabular}
\end{center}
\end{table}
\begin{table}[ht!] 
\footnotesize 
\begin{center}
\begin{tabular}{|c|c|c|c|c|c|c|}
\hline
$m_A$ (GeV) & \multicolumn{3}{|c|}{Cross-sections at $\sqrt{s}=13 \tev$} &  \multicolumn{3}{c|}{Cross-sections at $\sqrt{s}=14 \tev$} \\ 
\cline{2-7}
&  $\sigma_t$ (pb)& $\sigma_b$ (pb)& $\sigma_{\mathrm{tot}}$(pb)& $\sigma_t$ (pb)& $\sigma_b$ (pb)& $\sigma_{\mathrm{tot}}$(pb) \\
\hline
495  &6.68$\times10^{-2}$& $2.03\times10^{-6}$  &  6.69$\times10^{-2}$& $ 7.95\times10^{-2}$ &$2.42\times10^{-6}$ & $7.97\times10^{-2}$ \\
505  &6.26$\times10^{-2}$&  $1.85\times10^{-6}$ & 6.24$\times10^{-2}$ & $7.46\times10^{-2}$ &$2.22\times10^{-6}$& $7.45\times10^{-2}$  \\
515  &5.86$\times10^{-2}$&  $1.67\times10^{-6}$ & 5.86$\times10^{-2}$ & $6.97\times10^{-2}$ &$2.05\times10^{-6}$& $6.98\times10^{-2}$     \\
525&  5.48$\times10^{-2}$ & $1.52\times10^{-6}$ &5.49$\times10^{-2}$  & $6.55\times10^{-2}$&$ 1.89\times10^{-6}$ &  $6.54\times10^{-2}$    \\
535  &5.15$\times10^{-2}$& $1.38\times10^{-6}$  & 5.15$\times10^{-2}$ & $6.14\times10^{-2}$ &$1.76\times10^{-6}$ & $6.13\times10^{-2}$  \\
545  & 4.85$\times10^{-2}$&  $1.25\times10^{-6}$  &4.83$\times10^{-2}$ &$5.80\times10^{-2}$&$1.64\times10^{-6}$&  $5.77\times10^{-2}$  \\
555  &4.56$\times10^{-2}$&  $1.13\times10^{-6}$ &  4.54$\times10^{-2}$& $5.43\times10^{-2}$ & $ 1.52\times10^{-6}$& $5.43\times10^{-2}$  \\
565  &4.28$\times10^{-2}$& $1.02\times10^{-6}$ &4.28$\times10^{-2}$  & $5.14\times10^{-2}$&$1.41\times10^{-6}$ &  $5.12\times10^{-2}$   \\
575  &4.03$\times10^{-2}$&  $9.16\times10^{-7}$& 4.03$\times10^{-2}$ & $4.84\times10^{-2}$ &$1.32\times10^{-6}$ & $4.84\times10^{-2}$   \\
585  &3.81$\times10^{-2}$&  $8.27\times10^{-7}$&  3.79$\times10^{-2}$&  $4.57\times10^{-2}$&$ 1.22\times10^{-6}$ &  $4.57\times10^{-2}$  \\
595  &3.59$\times10^{-2}$ & $7.48\times10^{-7}$ & 3.59$\times10^{-2}$ & $4.32\times10^{-2}$&$1.14\times10^{-6}$& $4.29\times10^{-2}$ \\
605  &3.39$\times10^{-2}$&  $6.76\times10^{-7}$&3.38$\times10^{-2}$  &  $4.08\times10^{-2}$&$ 1.07\times10^{-6}$&  $4.06\times10^{-2}$ \\
615   &3.32$\times10^{-2}$& $6.10\times10^{-7}$  &3.19$\times10^{-2}$ &$3.85\times10^{-2}$ &$ 1.00\times10^{-6}$ & $3.84\times10^{-2}$  \\
625  & 3.04$\times10^{-2}$&  $5.51\times10^{-7}$ &3.03$\times10^{-2}$ &$3.64\times10^{-2}$ &$9.37\times10^{-7}$& $3.65\times10^{-2}$  \\
635  &2.87$\times10^{-2}$ &  $4.99\times10^{-7}$ &2.86$\times10^{-2}$ &$3.46\times10^{-2}$ &$8.76\times10^{-7}$ & $3.45\times10^{-2}$  \\
645  &2.72$\times10^{-2}$&   $4.51\times10^{-7}$&2.71$\times10^{-2}$ &$3.28\times10^{-2}$& $8.20\times10^{-7} $& $3.27\times10^{-2}$    \\
655  &2.58$\times10^{-2}$ & $4.07\times10^{-7}$  & 2.56$\times10^{-2}$&$3.11\times10^{-2}$& $7.73\times10^{-7}$& $3.11\times10^{-2}$  \\
665  &2.45$\times10^{-2}$&  $3.69\times10^{-7}$ &2.43$\times10^{-2}$ & $2.95\times10^{-2}$ &$7.24\times10^{-7}$  &$2.95\times10^{-2}$  \\
675  &2.31$\times10^{-2}$ & $3.36\times10^{-7}$  &2.31$\times10^{-2}$ &$2.79\times10^{-2} $&$6.79\times10^{-7}$ & $2.79\times10^{-2}$  \\
685 &2.19$\times10^{-2}$& $3.04\times10^{-7}$  &2.19$\times10^{-2}$ &$2.65\times10^{-2}$ &$6.39\times10^{-7} $& $2.65\times10^{-2}$ \\
695 & 2.09$\times10^{-2}$&  $2.74\times10^{-7}$ &2.08 $\times10^{-2}$ &$2.53\times10^{-2}$&$6.04\times10^{-7} $& $2.52\times10^{-2}$   \\
705 &1.98$\times10^{-2}$&  $2.47\times10^{-7}$ & 1.97$\times10^{-2}$ & $2.41\times10^{-2}$&$5.67\times10^{-7}$&  $2.39\times10^{-2}$  \\
715 &1.89$\times10^{-2}$ &  $2.22\times10^{-7}$ & 1.88$\times10^{-2}$ &$2.29\times10^{-2}$&$5.33\times10^{-7}$ & $2.27\times10^{-2}$ \\
725 &1.79$\times10^{-2}$ & $1.99\times10^{-7}$  &1.79$\times10^{-2}$  &$2.18\times10^{-2}$ &$5.03\times10^{-7}$ & $2.16\times10^{-2}$  \\
735 & 1.71$\times10^{-2}$&  $1.78\times10^{-7}$ & 1.69$\times10^{-2}$ &$ 2.08\times10^{-2}$&$4.73\times10^{-7}$ & $2.07\times10^{-2}$ \\
745 &1.62$\times10^{-2}$&  $1.60\times10^{-7}$ & 1.61$\times10^{-2}$ &$1.97\times10^{-2}$&$4.47\times10^{-7}$& $1.97\times10^{-2}$ \\
755& 1.54$\times10^{-2}$& $1.43\times10^{-7}$& 1.54$\times10^{-2}$  &$1.88\times10^{-2}$& $4.23\times10^{-7}$ &  $1.87\times10^{-2}$ \\
765 &1.47$\times10^{-2}$& $1.29\times10^{-7}$  &1.46$\times10^{-2}$  & $1.79\times10^{-2}$&$3.98\times10^{-7} $&  $1.79\times10^{-2}$  \\
775 &1.40$\times10^{-2}$ &  $1.16\times10^{-7}$ &1.40$\times10^{-2}$  &$1.71\times10^{-2}$&$ 3.77\times10^{-7}$&  $1.70\times10^{-2} $  \\
785 & 1.33$\times10^{-2}$& $1.04\times10^{-7}$ & 1.33$\times10^{-2}$ &$1.63\times10^{-2}$ &$3.58\times10^{-7}$ &  $1.63\times10^{-2}$  \\
795 & 1.28$\times10^{-2}$& $9.31\times10^{-8}$  & 1.27$\times10^{-2}$ &$1.56\times10^{-2}$&$3.38\times10^{-7}$&  $1.55\times10^{-2}$ \\
805 &1.22$\times10^{-2}$ & $8.35\times10^{-8}$  & 1.21$\times10^{-2}$  &$ 1.48\times10^{-2} $ &$3.21\times10^{-7}$ & $1.48\times10^{-2}$   \\
815 &1.16$\times10^{-2}$ &$7.55\times10^{-8}$   & 1.16$\times10^{-2}$  &$1.42\times10^{-2} $ &$ 3.04\times10^{-7}$ & $1.41\times10^{-2}$   \\
825 &1.11$\times10^{-2}$ &$6.82\times10^{-8}$   & 1.10$\times10^{-2}$  &$1.36\times10^{-2}$& $2.87\times10^{-7}$&  $1.35\times10^{-2}$   \\
835 &1.06$\times10^{-2}$ &$6.16 \times10^{-8}$ & 1.05$\times10^{-2}$  &$1.29\times10^{-2}$ &$2.73\times10^{-7}$ &  $1.29\times10^{-2}$  \\
845 &1.01$\times10^{-2}$ &  $5.57\times10^{-8}$ &  1.00$\times10^{-2}$ &$1.24\times10^{-2}$&$2.58\times10^{-7}$&  $1.24\times10^{-2}$  \\
855 &9.66$\times10^{-3}$ & $5.05\times10^{-8}$  &   9.62$\times10^{-3}$ &$1.18\times10^{-2}$&$2.45\times10^{-7} $&  $1.18\times10^{-2}$   \\
865 & 9.23$\times10^{-3}$&$4.55\times10^{-8}$   &  9.19$\times10^{-3}$ & $1.14\times10^{-2}$&$2.33\times10^{-7}$&  $1.13\times10^{-2}$   \\
875 &8.84$\times10^{-3}$& $4.13\times10^{-8}$  &8.78$\times10^{-3}$  &$1.09\times10^{-2}$ &$2.21\times10^{-7}$ &  $1.08\times10^{-2}$   \\
885&8.07$\times10^{-3}$ &$3.74\times10^{-8}$   &8.38$\times10^{-3}$    & $1.04\times10^{-2}$&$ 2.11\times10^{-7}$ &  $1.04\times10^{-2}$   \\
895 &7.71$\times10^{-3}$& $3.42\times10^{-8}$  &8.04$\times10^{-3}$   &$9.96\times10^{-3}$ &$2.00\times10^{-7}$ &  $ 9.93\times10^{-3}$  \\
905  &7.39$\times10^{-3}$&$3.10\times10^{-8}$   &7.71$\times10^{-3}$  &$9.56\times10^{-3}$& $1.89\times10^{-7}$&  $9.51\times10^{-3}$   \\
915 & $7.10\times10^{-3}$& $2.82\times10^{-8}$  & $7.37\times10^{-3}$  &$9.15\times10^{-3}$ &$1.81\times10^{-7}$ &  $ 9.13\times10^{-3}$   \\
925 &$6.79\times10^{-3}$&$2.58\times10^{-8}$ & $7.07 \times10^{-3}$  &$8.78\times10^{-3}$&$1.73\times10^{-7}$ &  $ 8.71\times10^{-3}$   \\
935 & 6.53$\times10^{-3}$&$2.35\times10^{-8}$   & $6.75\times10^{-3}$  &$8.42\times10^{-3}$&$1.65\times10^{-7}$ &  $8.38\times10^{-3}$  \\
945 &6.23$\times10^{-3}$&$2.15\times10^{-8}$ & $6.49\times10^{-3}$  &$8.03\times10^{-3}$ &$1.57\times10^{-7}$&  $8.04\times10^{-3}$  \\
955 &5.99$\times10^{-3}$ & $1.97\times10^{-8}$  & $6.21\times10^{-3}$  &$7.75\times10^{-3}$ &$1.49\times10^{-7}$ &  $7.69\times10^{-3}$  \\
965 &5.74$\times10^{-3}$&$1.80\times10^{-8}$ &   $5.95\times10^{-3}$ &$7.41\times10^{-3}$&$1.42\times10^{-7}$ &  $7.38\times10^{-3}$ \\
975 &5.50$\times10^{-3}$ & $1.65\times10^{-8}$  &  $5.46\times10^{-3}$  & $7.14\times10^{-3}$& $1.35\times10^{-7}$&  $7.08\times10^{-3}$ \\
985 &5.27$\times10^{-3}$ & $1.52\times10^{-8}$  & $5.27\times10^{-3}$ &$6.82\times10^{-3}$&$1.29\times10^{-7}$ &  $ 6.80\times10^{-3}$ \\
995&5.08$\times10^{-3}$ & $1.39\times10^{-8}$  & $5.08\times10^{-3}$ &$6.55\times10^{-3}$& $1.23\times10^{-7}$&  $6.53\times10^{-3}$  \\
1005&5.08$\times10^{-3}$ &$1.27\times10^{-8}$  & $5.04\times10^{-3}$  &$6.29\times10^{-3}$& $1.18\times10^{-7}$&  $6.25\times10^{-3}$   \\
\hline 
\end{tabular}
\end{center}
\end{table}

\clearpage
\pagebreak

{\footnotesize
\bibliographystyle{utphys}
\bibliography{biblio}
}

\end{document}